\documentclass[12pt]{article}

\usepackage{geometry}
\usepackage{fullpage}
\usepackage{graphicx}
\usepackage{pdflscape}
\usepackage{multirow}
\usepackage{ragged2e}

\usepackage{dcolumn}
\usepackage{booktabs,calc}
\usepackage{array}
\usepackage{paralist}
\usepackage{verbatim}
\usepackage{subfig}
\usepackage{setspace}
\usepackage{amsmath}
\usepackage{amssymb}
\usepackage{natbib}
\usepackage[linktocpage=true, colorlinks=true, linkcolor=blue, citecolor=black]{hyperref}
\usepackage{enumerate}
\usepackage{authblk}

\def\sym#1{\ifmmode^{#1}\else\(^{#1}\)\fi}

\begin{document}

	\title{Risk, Agricultural Production, and  Weather Index Insurance in Village India}
	
	\author[a]{Jeffrey D. Michler\thanks{Corresponding author email: jdmichler@email.arizona.edu. We greatly appreciate the helpful comments and suggestions from Philip Abbott, Hans Binswanger-Mkhize, Michael Roberts, and Neda Trifkovic along with those from seminar participants at the Econometric Society World Congress in Montreal, the Midwest International Economics Development Conference in Madison, the Agricultural and Applied Economics Association Annual Meeting in San Francisco, and the International Conference of Agricultural Economists in Milan. This paper and our thinking have been shaped by conversations with Steve Boucher, Garth Holloway, Mario Miranda, and Robert Townsend. We are solely responsible for any errors or misunderstandings.}}
	\author[b]{Frederi G. Viens}
	\author[c]{Gerald E. Shively}
	\affil[a]{\small \emph{University of Arizona, Tucson, USA}}
	\affil[b]{\small \emph{Michigan State University, East Lansing, USA}}
	\affil[c]{\small \emph{Purdue University, West Lafayette, USA}}
		
	\date{March 2021}
	\maketitle
	
\begin{center}\begin{abstract}
\noindent We investigate the sources of variability in agricultural production and their relative importance in the context of weather index insurance for smallholder farmers in India. Using parcel-level panel data, multilevel modeling, and Bayesian methods we measure how large a role seasonal variation in weather plays in explaining yield variance. Seasonal variation in weather accounts for 19-20 percent of total variance in crop yields. Motivated by this result, we derive pricing and payout schedules for actuarially fair index insurance. These calculations shed light on the low uptake rates of index insurance and provide direction for designing more suitable index insurance.
\end{abstract}\end{center}

{\small \noindent\emph{JEL Classification}: C11, D81, G22, O12, O13, Q16, Q12
	}

\newpage
\doublespacing
\section{Introduction}
Agricultural production is complex and risky. Weather is just one of several potential causes of yield variability. Other determinants include the quantity and quality of inputs, the agronomic characteristics of farmed parcels, the inherent or learned abilities of farmers, the policy environments in which farmers operate, and changes in technology \citep{HardakerEtAl97}. In this paper we use parcel-level panel data from India to measure the sources of variability in agricultural production and assess their relative importance. Ascertaining just how large a role weather plays in determining yields is vital to estimating the projected effects of climate change on a variety of outcomes, including aggregate economic activity \citep{SchlenkerEtAl06, DeschenesGreenstone07, BurkeEtAl15}. Understanding weather induced variation in yields is also essential when designing and implementing weather index insurance, an increasingly popular micro-level risk management strategy in developing countries. However, recent pilot projects and randomized control trials in Asia have found limited farmer uptake of index insurance \citep{HazellEtAl10, GineEtAl12, ColeEtAl13, CaiEtAl15} Using a multilevel modeling approach and Bayesian estimation we find that a relatively small fraction of the variability in yields can be attributed to seasonal variation in weather. While weather variability does not contribute much to yield variability, on average, our Bayesian methodology reveals the presence of infrequent but potentially costly extreme weather events. This result motivates our reassessment of the pricing of rainfall insurance and may partly explain the low uptake of weather index insurance by farmers.

The goal of weather index insurance is to assist farmers in managing covariate risk. In developing country agriculture, covariate risk is particularly difficult to informally insure against since, by its very nature, it affects neighboring households and limits the effectiveness of traditional risk sharing mechanisms. Several early studies looked at the role of weather risk on agricultural production and insurance in India. \cite{Townsend94} found that household consumption moves with village-level consumption and is not affected much by idiosyncratic shocks, which households cover via borrowing, gifts, and asset sales. He concludes that while households can insure against individual risks, covariate risk remains a problem. Along similar lines, \cite{RosenzweigBinswanger93} focus on the role of weather risk on production and asset portfolios. They find that village-level rainfall variables explain only a small proportion of household-level profit variability. And, similar to \cite{Townsend94}, they find households insuring against non-covariate risk. In part, the recognition that households are less successful in insuring against covariate weather risk than in insuring against non-covariate risk motivates recent attempts to design and implement weather-based index insurance for poor farmers.

This apparent need for weather index insurance has been met by surprisingly little uptake of the product by farmers in India \citep{GineEtAl08,GineEtAl12}. There are many potential barriers to uptake of insurance, including trust in the insurance company, household liquidity constraints, and lack of financial literacy of household decision makers \citep{ColeEtAl13}.\footnote{For example, \cite{BanerjeeEtAl14} find no demand among Indian households for health insurance bundled with microfinance, even among those for whom there was clear value. They attribute this low uptake to poor understanding of the insurance product and poor support for enrolling by insurance underwriters.} In addition to these consumer-centric explanations for low uptake, flaws in the product itself may be a cause, specifically the imperfect correlation between crop yield and farm profit, the outcome of ultimate interest to farmers \citep{Binswanger12}. This uninsured exposure, or basis risk, can be a significant deterrent to purchasing insurance by reducing the utility gain for households. There are two potential sources of basis risk. One is the loss caused by an event not measured by the index. This could include, for example, low temperatures that retard crop growth or high winds that cause crop damage, when the insurance index is based on rainfall. Another potential source of basis risk is low correlation between the index and covered losses. This can occur, for example, when measurement stations that are used to trigger payouts are spatially distant from insured crops and provide poor congruence with local conditions. A less common example is potentially insured crops that are sufficiently resilient to the phenomena which forms the basis for the insurance, such as drought-tolerant varieties covered by rainfall-based insurance or pest-resistant cotton to pest infestation \citep{Liu13}. In this paper we focus on these production-oriented sources of basis risk and empirically estimate the amount of variability in yield coming from seasonal weather variability.

The most common type of weather index insurance in India is rainfall insurance \citep{BarnettMahul07, GineEtAl07, AkterEtAl09}. However, surprisingly little empirical research has been conducted to confirm the underlying relationship between rainfall or seasonal weather variability and yield variability. \cite{BarnettMahul07} suggest that rainfall variability accounts for $50$ percent of yield variability but do not support this figure with data or a source. \cite{GineEtAl12} assert that as much as 90 percent of variation in Indian crop production is driven by rainfall volatility, citing \cite{Parchure02} as their source. But Parchure relies on a 1976 report by the Indian National Commission on Agriculture (NCA), the details of which are somewhat opaque.\footnote{According to \cite{Parchure02}, the $90$ percent variation is actually for cotton and groundnuts while variation in yield due to rainfall variability is $45$ and $47$ percent for wheat and barley, respectively. \cite{Parchure02} actually cites the 1976 NCA report as stating ``that rainfall variations accounted for $50\%$ of the variability in agricultural yields.'' This may be where \cite{BarnettMahul07} get their number. Examining the report, specifically pages 47-8, these numbers were obtained from a linear regression ``with yield as the dependent variable and total rainfall during the five crop growth phases as the independent variable'' \citep{NCA76}. The $50$ percent number is apparently the R$^2$ on the regression.} Even less data exist on the relationship between yield variability and other measures of covariate risk, such as temperatures, wind speed, or the occurrence of natural disasters. Finally, to our knowledge, no data exist on covariate risk's share in explaining variance in crop yield, which is where we focus our attention. Previous studies that have attempted such measurements were constrained by data, econometric techniques, and computing power so that they were unable to net out measurement error and fully describe the sources of output variability. We attempt to rectify these shortcomings and fill this research gap.

We examine the different sources of yield variability using a multilevel/hierarchical regression framework. This approach more fully accounts for the covariance structure of the data than a standard regression framework and allows us to control for inputs at the parcel-level and also to isolate the amount of yield variance due to parcel-level effects, household-level effects, seasonal weather effects, village-level effects, and time. Using Bayesian methods, we draw the underlying distribution of the random error term corresponding to different sources of variability, thereby providing a quantitative measure of potentially insurable risk. Bayesian methods are particularly useful because the underlying distribution for several of the error terms turn out to be highly skewed and non-normal. Considering all sources of yield variance, we find that seasonal variation in weather accounts for 19-20 percent of total variance in agricultural production.

While we conclude that weather variability accounts for a much smaller share of yield variability than has been previously argued, our results should not be interpreted as implying that smallholder farmers do not need weather insurance. Rather, what they need is better insurance - specifically, insurance that reduces the basis risk that results from a low correlation between the index and crop loss. Furthermore, skewness in the distribution for seasonal weather variability highlights the need for access to affordable risk-management tools. Indeed this skewness is a sign that, despite the low share of seasonal weather variation in yield variability, extreme weather events, while rare, are nevertheless potentially costly.

Motivated by these findings, in the second part of the paper we shift our focus to the closely related concern of proper pricing of index insurance contracts. We provide a brief quantitative reassessment of the index insurance contracts studied by \cite{GineEtAl07}, \cite{GineEtAl12}, and \cite{ColeEtAl13}. We conclude that these contracts are overpriced. This further underscores that flaws in the product are a proximate cause of low uptake and that farmers may rationally prefer to focus their risk management choices on minimizing other sources of risk.

To tie our results as closely as possible to the issue of weather risk in agricultural production, we use the International Crops Research Institute for the Semi-Arid Tropics (ICRISAT) household survey data from the same Indian villages studied by \cite{Townsend94} and \cite{RosenzweigBinswanger93}. However, as a further extension, we use an expanded data set that covers 24 newly added villages. In our analysis we use a panel of 10 cropping seasons from 2009 through 2013. This time horizon has both advantages and disadvantages. The primary benefit is that over a short time frame parcel and household characteristics, along with policy and technology, are unlikely to have significantly changed. By controlling for these ``fixed'' effects, along with parcel-level variable inputs in each time period, we can measure the variance in crop output resulting from weather risk. The primary shortcoming of our dataset is that the 10 seasons of production we observe contains only one large-scale weather-related natural disasters: the 2009 drought. This means that our measure of weather's share in total yield variability may be an underestimate.\footnote{That said, India has experienced only two large-scale droughts in the last 25 years (2002 and 2009). The Indian Meteorological Department considers a drought to have occurred when seasonal rainfall deficiency falls is $26$ percent below average rainfall.} Nevertheless, our analysis sheds new empirical light on the potential value of weather insurance during typical production periods. Our five year time horizon is also likely to mimic the typical time scale under which farmers in village India make investment decisions.

By using a multilevel approach to measuring the roles of idiosyncratic and covariate risk in agricultural production in village India we contribute to three separate streams of research. The first is the literature focused on production risk and insurance in the developing world. Our aim is to connect two strands of this research. The first, originating with \cite{Townsend94}, \cite{RosenzweigBinswanger93}, \cite{RosenzweigWolpin93}, and \cite{Kochar99} has focused on the theoretical and empirical relationship between production, risk, and insurance. The second focuses on interventions to help insure production against risk \citep{BarnettMahul07, GineEtAl07, SkeesEtAl07, GineEtAl08, AkterEtAl09, GauravEtAl11, ClarkeEtAl12a, ChantaratEtAl13, ColeEtAl13}. We empirically estimate the role different sources of variability play in determining the mean and variance of crop output. We then use this information to inform the value and pricing of index insurance products.

Our second contribution is to the nascent economics literature that applies multilevel models to household data. Although we introduce no new methods, per se, in the paper, we do contribute a methodological innovation by expanding the use, understanding, and adaptability of multilevel modeling, Bayesian inference, and Gibbs sampling, all in the context of household data.

The third body of literature to which we contribute is the empirical estimation of production functions. Until recently it has been computationally difficult to estimate production functions with multiple nested levels of data. While panel data fixed effects methods are equivalent to a multilevel model with a single level, few alternatives exist for models with a large number of nested levels. Production data are often collected at the farm or factory level, which suggests at minimum a two-level model.  Adding additional levels to account for and distinguish among temporal and/or spatial variation can provide more efficient estimation of production by more accurately modeling the data generating process. We contribute to this strand of literature by comparing production function parameter estimates obtained by standard ordinary least squares to multilevel regression estimates obtained by maximum likelihood and Bayesian estimation techniques.

\section{Data}\label{sec:data}
To conduct our empirical analysis, we use household data from villages in India. These data were collected as part of the Village Level Studies/Village Dynamics Study of South Asia \citep{VDSA13}. The data set combines high and low frequency household data from 30 Indian villages. The villages include the three studied by \cite{Townsend94} and the ten studied by \cite{RosenzweigBinswanger93}. However, while much of the previous research has relied on the low frequency data, we utilize a newly available high frequency data set covering the years 2009 through 2013. These data include monthly household observations on input purchases and labor expenditure for on-farm activities and crop production. The added value of a high frequency data set is that, with multiple crops on multiple parcels for multiple seasons in a single year, it provides much more detailed and accurate farm production information than the low frequency data.

We utilize monthly parcel-level data aggregated to the seasonal level. We focus on five crops: paddy rice, sorghum, wheat, maize, and cotton. Together these crops account for 60 percent of total crop observations in the VDSA and cover 78 percent of the total parcel-level observations.\footnote{With the exception of pigeon pea, a perennial crop, the five crops used in our analysis are the five most frequently observed crops in the data set. We include cotton instead of pigeon pea for two reasons. First, pigeon pea is a perennial crop and therefore may be treated by farmers differently than annual crops when considering insurance. Second, cotton is one of the crops considered by \cite{ClarkeEtAl12b} in their product design and ratemaking for insurance contracts in Gujarat. Thus, despite it being a non-food crop, we include it to bring our analysis more closely in line with existing literature.} This provides us with 11,942 parcel-level observations. Rice is the most common crop, accounting for 46 percent of total observations. Next most common is sorghum, accounting for 23 percent of observations. Wheat constitutes 14 percent of observations while maize and cotton account for the remaining 9 and 8 percent of observations respectively.

The locations of VDSA villages provides us with a great deal of heterogeneity in climate, weather, crop choice, and cultivation practices. Statistics describing these data, by crop, are presented in Table~\ref{tab:cropdisc}. Twenty-seven villages cultivate at least two crops, generally a single crop in each of the two growing seasons. Only two villages cultivate all five crops. Cotton is the most input intensive crop, using more labor, fertilizer, mechanization, and pesticide than any other crop. Sorghum is the least labor intensive while wheat uses the least fertilizer, mechanization, and pesticide. In total, our 11,942 parcel-level observations come from 5,100 unique parcels operated by 1,079 distinct households, in 30 villages, farming across 10 seasons. We exploit this nested data structure in our empirical analysis.

One final source of variation in the data structure, which plays an important role in our analysis, is the number of time observations for each crop. Three crops (rice, wheat, and maize) are grown in both the \emph{Kharif} monsoon season and the post-monsoon \emph{Rabi} season. Sorghum is only grown in \emph{Rabi} while cotton is only grown in \emph{Kharif}. The limited time series for these crops may affect our estimation of seasonality's effects on yield variability, albeit in an unknown way.

\section{Econometric Framework}\label{sec:empirical}

\subsection{Ordinary Least Squares}
We begin by estimating a simple linear regression for yield. Let $y_{is}$ denote the log of yield for parcel $i$ in season $s$. We estimate

\begin{equation}
y_{is} = X_{is} \beta + \alpha_s + \epsilon_{is} \label{eq:OLS}
\end{equation}

\noindent where $X_{is}$ is a matrix of data and $\beta$ is a vector of regression coefficients associated with various crops. In order to account for the spatial dimension and allow our proxy for weather to vary across both location and time, the seasonal fixed effect $\alpha_s$ is an indicator for each time period $(t=10)$ in each village $(v=30)$. We assume the error term is $\epsilon_{is} \sim \mathcal{N}(0,\sigma^2)$ so that $y_{is} \sim \mathcal{N}(X_{is} \beta + \alpha_s,\sigma^2)$, where $\beta$ and $\sigma^2$ are regression and variance parameters which are season independent while $\alpha_s$ depends on the season. 

We note several drawbacks associated with this linear estimation of the production function. First is that it precludes one from discerning the role of weather in yield variability. Equation~\eqref{eq:OLS} can estimate the impact of parcel-level inputs on yield at each point in time as well as the impact of a seasonal weather dummy on yield. But because these variables impact mean yield, the specification does not allow us to measure the share of yield variability $(\sigma^2)$ represented by seasonal variability in weather. A second drawback is that OLS limits our ability to control for additional clustered effects, such as parcel, household, village, or temporal effects.  While changes in season clearly impact the effectiveness of parcel-level inputs, equally relevant effects may exist at these other levels. Some households may be more efficient in their application of labor compared to others, while some parcels may be of better quality, resulting in less need for, say, fertilizer. Some households may live in villages in states with favorable agricultural policy resulting in better access to inputs. Finally, while we do not expect much technological change between 2009 and 2013, the quality of inputs continues to evolve, which may impact yields over time. Even if it were computationally feasible to estimate season-specific, household-specific, and parcel-specific parameters using OLS, such grouped data would violate the assumption of independence for all data \citep{CorradoFingleton12}. A third drawback, and perhaps the most important for us, is that OLS requires the variance terms to be normally distributed. While in many applications this assumption is appropriate, we should not expect weather events to be normally distributed. If time-specific effects have a highly skewed distribution, OLS will produce biased parameter estimates and estimates of risk severity may ignore extreme events.

\subsection{The Multilevel Model}
A multilevel or hierarchical modeling strategy addresses the first two drawbacks associated with the standard linear approach to estimation.\footnote{\cite{GelmanHill07} provide an introduction to multilevel analysis.} First, multilevel models offer a natural way to assess the role of seasonal changes in weather on variation in yields by explicitly modeling the variance, not just the mean of the data. This allows us to measure the different sources of variance in yields. In our case, a multilevel approach also allows us to disaggregate total variance in yields into its multiple sources, so as to measure the relative contribution of seasonal weather risk in production. Second, a multilevel approach allows us to control for each grouping of the data without adding to the computational burden and without violating independence assumptions.

For expository purposes we start with an illustrative example of a simple two-level model in which realizations of yields are grouped within seasons. Let $y_{ns}$ denote the log of an observed yield, $n$, realized in season $s$. We estimate

\begin{subequations}
	\begin{align}
	y_{ns} &= X_{n} \beta + \alpha_{s} + \epsilon_{ns} \label{eq:2level1} \\
	\alpha_{s} &= \mu + \nu_{s} \label{eq:2level2}
	\end{align}
\end{subequations}

\noindent where $X_{n}$ and $\beta$ are as previously defined, and $\alpha_{s}$ is a seasonal effect that is a function of an overall mean, $\mu$, and a random disturbance term, $\nu_s$. We assume that $\epsilon_{ns} \sim \mathcal{N} (0, \sigma_{\epsilon}^{2} )$, $\nu_{s} \sim \mathcal{N} (0, \sigma_{\nu}^{2} )$, and $\epsilon_{ns}$ is independent of $\nu_{s}$.

In order to make our parameter of interest explicit, we can rewrite equations~\eqref{eq:2level1} and \eqref{eq:2level2} in terms of a probability distribution so that

\begin{equation}
y_{ns} \sim \mathcal{N}(X_{ns} \beta + \mu,u_{n}), \label{eq:prob1}
\end{equation}

\noindent with $u_{n} \equiv \sigma_{\nu}^{2} + \sigma_{\epsilon}^{2}$. The above distribution is obtained by substituting \eqref{eq:2level2} into \eqref{eq:2level1} and using the independence of $\epsilon_{ns}$ and $\nu_{s}$.  Defining the regression equation in this way highlights the very specific dispersion structure of the residual, which is where our interest lies. It also allows us to easily define the intraclass correlation coefficient (ICC),

\begin{equation}
\rho = \frac{\sigma_{\nu}^{2}}{\sigma_{\nu}^{2} + \sigma_{\epsilon}^{2}}, \label{eq:icc}
\end{equation}

\noindent which is similar to the proportion of explained variance in an OLS regression.

The value of a multilevel model becomes obvious as we add additional levels. In our analysis we can view each observation on yield, $y_n$, as coming from a parcel group $i$; each parcel group as being nested within a household, $h$; each household as observed during a specific season, $s$; each season occurring in a unique village, $v$; and each village existing at a given time, $t$.\footnote{Note that $i$, $h$, $s$, $v$, and $t$ can be understood as functions of $n$ so that each unique data point corresponds to the index $n$, each data point can be identified with a unique parcel $i$, each parcel can be identified with a unique household $h$, each household experiences a unique season $s$, each season occurs in a unique village $v$, and each village 	exists at unique time $t$.} We can write the multilevel model as:

\begin{subequations}
	\begin{align}
	&\textbf{Level 0 } \mathrm{(yields}, n=11,942): y_{n} = X_{n} \beta + \alpha_{ihsvt} + \epsilon_{n} \label{eq:level0} \\
	&\textbf{Level 1 } \mathrm{(parcels}, i=5,100): \alpha_{ihsvt} = \alpha_{hsvt} + \nu_{ihsvt} \label{eq:level1} \\
	&\textbf{Level 2 } \mathrm{(households}, h=1,079): \alpha_{hsvt} = \alpha_{svt} + \nu_{hsvt} \label{eq:level2} \\
	&\textbf{Level 3 } \mathrm{(seasons}, s=240): \alpha_{svt}  = \alpha_{vt} + \nu_{svt} \label{eq:level3} \\
	&\textbf{Level 4 } \mathrm{(villages}, v=30): \alpha_{vt}  = \alpha_{t} + \nu_{vt} \label{eq:level4} \\
	&\textbf{Level 5 } \mathrm{(time}, t=10): \alpha_{t}  = \mu + \nu_{t} \label{eq:level5} 
	\end{align}
\end{subequations}

\noindent where again $X_{n}$ is a matrix of input data and $\beta$ is a vector of regression coefficients associated with the various crops.

At level 0 the model estimates the log of yield as a function of inputs, similar to that in equation~\eqref{eq:OLS}, with yields being a function of a specific parcel $i$ and an idiosyncratic error term $\epsilon_{n} \sim \mathcal{N}(0,\sigma^{2})$, where $\sigma^{2}$ is a constant variance parameter which we assume does not depend on $i$, $h$, $s$, $v$, or $t$. Each $n$ observation comes from a parcel cluster $i$ which we assign a unique intercept, $\alpha_{ihsvt}$. This parcel-level intercept allows the relationship between inputs and yield to differ across parcels depending on parcel-level characteristics. While some parcel characteristics can be observed, many are difficult to measure or costly to observe. Such characteristics include soil micro-nutrients, grade, and aeration or composition. By including a unique intercept term for each parcel we can control for these parcel characteristics.

Level 1 of the model groups parcels within households. Here parcel intercepts, $\alpha_{ihsvt}$, are a function of household characteristics, $\alpha_{hsvt}$, and a random disturbance term, $\nu_{ihsvt} \sim \mathcal{N}(0,\sigma_{1}^{2})$, where $\sigma_{1}^{2}$ is a constant variance parameter. We assume that the $\nu_{ihsvt}$ terms are independent of each other and that the vectors $\epsilon_{n}$ and $\nu_{ihsvt}$ are independent. The household-level intercept allows variation in parcel-level production to be dependent on household characteristics. In most applications, the analyst attempts to control for unobserved household ability through proxy variables such as age or education. The multilevel approach allows us to control for any unobserved household-level characteristics by assigning each household a unique intercept term without the need to rely on proxies. The use of a single disturbance term for all data points $n$ corresponding to a given parcel group further enhances this control by imposing a covariance structure which is consistent with variation at the parcel group level.

Level 2 of the model groups households within seasons. Here household-level intercepts, $\alpha_{hsvt}$, are a function of season, $\alpha_{svt}$, and a random disturbance term, $\nu_{hsvt} \sim \mathcal{N}(0,\sigma_{2}^{2})$,  where $\sigma_{2}^{2}$ is a constant variance parameter.\footnote{As in the case of the level 1 disturbance terms, the terms $\nu_{hsvt}$ are assumed to be independent of each other. The corresponding vector $\nu_{hsvt}$ is assumed to be independent of $\epsilon_{n}$ and $\nu_{ihsvt}$. The same group-level independence assumptions hold for all levels.} The season-level intercept allows variation in household-level efficiency to depend on seasonal weather events. While household ability is often viewed as time invariant, it will be time dependent if household ability is diminished or enhanced by changes in weather \citep{Kochar99}. Households experienced in dealing with droughts may find their ability diminished by flooding or cyclones. \emph{A priori}, there is no reason to assume that seasonality or changes in weather have a constant or stationary effect on household characteristics. By allowing household-level intercepts to vary across season, we are relaxing the assumption that household characteristics are either time invariant or affected by weather in the same way each season.

Level 3 of the model groups seasons within villages. Here season-level intercepts, $\alpha_{svt}$, are a function of a village, $\alpha_{vt}$, and a random disturbance term, $\nu_{svt} \sim \mathcal{N}(0,\sigma_{3}^{2})$,  where $\sigma_{3}^{2}$ is a constant variance parameter. The village-level intercept allows for season-on-season variation of average yields within each village. By making seasonality a function of both village and time, we are able to account for both the spatial and temporal nature of weather. This allows our model to account for the fact that not all crops are grown in each village and in each time period.

Level 4 of the model groups villages within time. Here village-level intercepts, $\alpha_{vt}$, are a function of time, $\alpha_{t}$, and a random disturbance term, $\nu_{vt} \sim \mathcal{N}(0,\sigma_{4}^{2})$,  where $\sigma_{4}^{2}$ is a constant variance parameter. The time-level intercept allows variation in village-level effects to depend on time. While our data cover only five years, we cannot rule out changes in policy, technology, or both during this period. In India, the introduction of new technologies and the implementation of new policies frequently occur at the state level and thus will not uniformly effect all villages. By allowing our village-level intercepts to vary across time we can control for unobservable technological and political changes.

Level 5 of the model defines the temporal intercepts as a function of an overall mean, $\mu$, and a random disturbance term, $\nu_t \sim \mathcal{N}(0,\sigma_{5}^{2})$, where $\sigma_{5}^{2}$ is a constant variance parameter. Having controlled for all other potentially relevant sources of yield variability, we interpret seasonal variation as coming solely from weather events.

The model represented above by equations~\eqref{eq:level0}-\eqref{eq:level5} and its dependence/independence structures can be summarized as a single model in terms of a probability distribution with a special error structure which is a sum of independent disturbance terms with a nested dependence on indexes:

\begin{equation}
y_{n} \sim \mathcal{N}(X_{n} \beta + \mu,u_{n}), \label{eq:single2}
\end{equation}

\noindent where $u_{n}$ is a specific covariance matrix which is the sum of six covariance matrices corresponding to the disturbance vectors $\epsilon_n$, $\nu_{ihsvt}$, $\nu_{hsvt}$, $\nu_{svt}$, $\nu_{vt}$, and $\nu_t$ from each level. Considered as a single error term, this error structure is non-trivial, though the expression shows how it works intuitively. Equation~\eqref{eq:single2} makes clear that our parameters of interest are not the additive non-interacting scale terms (the $\alpha$'s) but the components of the error term $(\sigma^{2}$, $\sigma_{1}^{2}$, $\sigma_{2}^{2}$, $\sigma_{3}^{2}$, $\sigma_{4}^{2}$, $\sigma_{5}^{2})$. In particular, our model, whether represented by equations~\eqref{eq:level0}-\eqref{eq:level5} or by equation~\eqref{eq:single2}, shows that the main goal in uncertainty quantification is to estimate the disturbance terms. Indeed, this is the key to evaluating the share of variance in yield corresponding to each level in the hierarchy.  

The ICC for the five-level model is the percentage of the total variance that is explained by the variance within clusters of groups. So, the correlation between realizations of yield within the same parcel is

\begin{equation}
\rho(parcel) = \frac{\sigma_{1}^{2}}{\sum_{i=0}^{5} \left( \sigma_{i}^{2}\right) }. \label{eq:icc1}
\end{equation}

\noindent The correlation between realizations of yield within the same household is

\begin{equation}
\rho( household ) = \frac{\sigma_{2}^{2} + \sigma_{1}^{2}}{\sum_{i=0}^{5} \left( \sigma_{i}^{2}\right) }. \label{eq:icc2}
\end{equation}

\noindent The correlation between realizations of yield within the same season is

\begin{equation}
\rho( season ) = \frac{\sigma_{3}^{2} + \sigma_{2}^{2} + \sigma_{1}^{2}}{\sum_{i=0}^{5} \left( \sigma_{i}^{2}\right) }. \label{eq:icc3}
\end{equation}

\noindent The correlation between realizations of yield within the same village is

\begin{equation}
\rho( village ) = \frac{\sigma_{4}^{2} + \sigma_{3}^{2} + \sigma_{2}^{2} + \sigma_{1}^{2}}{\sum_{i=0}^{5} \left( \sigma_{i}^{2}\right) }. \label{eq:icc4}
\end{equation}

\noindent And, the correlation between realizations of yield during the same time period is

\begin{equation}
\rho( time ) = \frac{\sigma_{5}^{2} + \sigma_{4}^{2} + \sigma_{3}^{2} + \sigma_{2}^{2} + \sigma_{1}^{2}}{\sum_{i=0}^{5} \left( \sigma_{i}^{2}\right) }. \label{eq:icc5}
\end{equation}

\noindent By construction the ICC increases as we move to higher levels of aggregation. Thus, we also calculate each level's contribution to total variance in the model. This is simply the variance at each level divided by $u_{n}$, the total variance in yields.

\subsection{Bayesian Estimation of the Multilevel Model}
While multilevel models address the first two drawbacks of OLS estimation, they still rely on the standard, though potentially unsupported, assumption of normality of the disturbance term at each level. We can verify if the normality assumption placed on each of the random disturbance terms is reasonable using a likelihood ratio (LR) test. We visualize the results of these tests using zeta profile plots, which plot the sensitivity of the model fit to changes in values of particular parameters. While these plots are not equivalent to drawing the underlying distributions of the estimators, they represent a similar idea and can be interpreted as representing the underlying distributions. First, we estimate the model in equations~\eqref{eq:level0}-\eqref{eq:level5} using maximum likelihood. Then we hold a single parameter fixed and vary the other parameters, assessing the fit of each new iteration compared to the globally optimal fit using the LR as our comparison statistic. We then apply a signed square root transformation to the LR statistic and plot the absolute value of the resulting function, $| \zeta |$, in comparison to the estimated parameter values. The zeta profile plots resulting from the model fit to equations~\eqref{eq:level0}-\eqref{eq:level5} are presented in Figure~\ref{fig:zeta}.\footnote{Given the large number of parameters ($20$ coefficients and $6$ disturbance terms) we prefer the graphical representation of the outcomes of the LR tests. Numerical results of the tests are available from the authors upon request.} Parameters with underlying normal distributions have straight line zeta profile plots. When this is the case these parameters provide a good approximation to the normal distribution and standard confidence intervals can be used for inference. When zeta profile plot lines are not straight, the normal distribution is a poor approximation of the underlying distribution. For parameter estimates on data, a non-normal distribution simply requires an adjustment of the relevant test statistics. But, if the underlying distributions of our disturbance parameters are non-normal, we are faced with a violation of the normality assumptions of the error term. In Figure~\ref{fig:zeta}, zeta profile plots for the data, the parcel disturbance term $(\sigma_1)$, and the idiosyncratic disturbance term $(\sigma)$ represent good approximations of normality. The zeta profile plot for the weather disturbance term $(\sigma_3)$ and the village disturbance term $(\sigma_4)$ are slightly skewed. The zeta profile plot for the household disturbance term $(\sigma_2)$ and the time disturbance term $(\sigma_5)$ are highly skewed and cannot be assumed to follow a normal distribution. Thus, we cannot assume normality of the individual error terms nor the composite error term. Since we do not, \emph{a priori}, know the distribution of $\epsilon_n$, maximum likelihood methods cannot be used to reliably estimate our model.

We address this drawback to classical estimation of multilevel models by adopting a Bayesian framework and using Markov Chain Monte Carlo (MCMC) methods to obtain posterior estimates. For estimation of posterior distributions we use the Gibbs sampler, which iteratively constructs a sequence of samples from the univariate random values of each response variable, and of each model parameter, conditional on all other parameters and variables. This method allows us to compute all features of the marginal and joint distributions because the marginal samples are iteratively fed back into the conditional posterior densities of all other parameters and variables for each sampling. This allows us to calculate unbiased point estimates and confidence intervals for all variables without recourse to normality assumptions.

We define the Bayesian estimator using our illustrative two-level model and then provide parameter definitions for the specific model defined by equations~\eqref{eq:level0}-\eqref{eq:level5}.\footnote{\cite{CameronTrivedi05} and \cite{HanmakerKlugist11} outline Bayesian estimation of multilevel models.} For a general two level model, let $y_{km}$ be a unique observation $k = 1, ..., K$ from the group $m = 1, ..., M$. Within each group $m$, the data are distributed according to a particular distribution $G$ with parameter $\gamma$ such that $y_{km} \sim G(\gamma_m)$. We assume that parameter $\gamma_m$ comes from a distribution $L$ with parameter $\lambda$ such that $\gamma_{m} \sim L(\lambda)$. Finally, we assign a distribution to the hyperparameter $\lambda$ so that $\lambda \sim Q(a,b)$, where $a$ is the mean and $b$ is the variance of the distribution. The joint posterior distribution of all unknown parameters is derived using Bayes' theorem:

\begin{equation}
p(\lambda,\gamma | y) \propto p(y | \gamma, \lambda) p(\gamma | \lambda)p(\lambda)
\end{equation}

\noindent where $y = (y_{11}, ..., y_{1K}, ..., y_{M1}, ..., y_{MK})$ is all the data, $\gamma = (\gamma_1, ..., \gamma_M)$ is the group level parameter, and  $\lambda$ is the population parameter. The density of the data is obtained through

\begin{equation}
p(y | \gamma, \lambda) = \prod_{m=1}^{M} \prod_{k=1}^{K} p(y_{km} | \gamma_m, \lambda). \label{eq:dprod}
\end{equation}

\noindent This is an independence assumption, and the individual density, $p(y_{km} | \gamma_m, \lambda)$, is assumed to be known. Again by the independence assumption, the prior for the group level effect is assumed to be

\begin{equation}
p(\gamma | \lambda) = \prod_{m=1}^{M} p(\gamma_m | \lambda). \label{eq:sprod}
\end{equation}

\noindent From the joint posterior distribution we can derive the conditional posterior distributions.\footnote{In terms of the independence assumptions made for the MLE multilevel model, assumption~\eqref{eq:dprod} corresponds exactly to assuming that the error terms $(\epsilon_n)$ are independent of each other while assumption~\eqref{eq:sprod} corresponds exactly to assuming that the elements in each group-level $\nu$ term are independent of each other. The Bayesian multilevel setup in which each group-level parameter is assumed to have its own prior distribution is consistent with assuming in the MLE multilevel model that $\epsilon_n$ and the $\nu$'s are independent.}

The Bayesian estimator for our multilevel model given by equations~\eqref{eq:level1}-\eqref{eq:level4} can be stated by redefining the model in probability terms:

\begin{subequations}
	\begin{align}
	\textbf{Level 0}:& \ \ y_{n} \sim \mathcal{N} \left( X_{n} \beta +  \alpha_{ihsvt}, \sigma^{2} \right) \label{eq:Blevel0} \\
	\textbf{Level 1}:& \ \ \alpha_{ihsvt} \sim \mathcal{N} \left( \alpha_{hsvt}, \sigma_{1}^{2} \right) \label{eq:Blevel1}  \\
	\textbf{Level 2}:& \ \ \alpha_{hsvt} \sim \mathcal{N} \left( \alpha_{svt}, \sigma_{2}^{2} \right) \label{eq:Blevel2}  \\
	\textbf{Level 3}:& \ \ \alpha_{svt} \sim \mathcal{N} \left( \alpha_{vt}, \sigma_{3}^{2} \right) \label{eq:Blevel3} \\
	\textbf{Level 4}:& \ \ \alpha_{vt} \sim \mathcal{N} \left( \alpha_{t}, \sigma_{4}^{2} \right) \label{eq:Blevel4} \\
	\textbf{Level 5}:& \ \ \alpha_{t} \sim \mathcal{N} \left( \mu, \sigma_{5}^{2} \right). \label{eq:Blevel5}
	\end{align}
\end{subequations}

\noindent The hyperpriors, or the prior distributions of the hyperparameters,  are defined as:

\begin{eqnarray}
\textbf{Hyperpriors}:
&  &\mu \sim \mathcal{N} \left( a, b \right) \nonumber \\
&  &\sigma_{4}^{2} \sim IG \left( d_4,g_4 \right) \nonumber \\
&  &\sigma_{3}^{2} \sim IG \left( d_3,g_3 \right) \\
&  &\sigma_{2}^{2} \sim IG \left( d_2,g_2 \right) \nonumber \\
&  &\sigma_{1}^{2} \sim IG \left( d_1,g_1 \right) \nonumber \\
&  &\sigma_{0}^{2} \sim IG \left( d_0,g_0 \right) \nonumber
\end{eqnarray}

\noindent where $IG$ is the inverse gamma distribution. The assumption that the priors are distributed inverse gamma is a common assumption in Bayesian econometrics in order to facilitate the mathematical analysis via the Gibbs sampler. We select values for the distributions to ensure that priors are uninformative. We then use a burn-in period of 5,000 iterations and an additional 5,000 iterations to ensure convergence of the iterative simulations and sufficient mixing of the Gibbs sampler.

\section{Econometric Results}\label{sec:discussion}
We present the results from a large complement of regressions in Tables~\ref{tab:reg} -~\ref{tab:crops}. Table~\ref{tab:reg} presents a trio of results from fitting the production function using OLS, MLE, and Bayes. Table~\ref{tab:ICC} presents estimated variances, ICCs, and variance shares from the MLE and Bayesian versions of the production function. Table~\ref{tab:crops} presents both MLE and Bayesian estimation results from crop-by-crop production functions.\footnote{In the interests of parsimony we focus on the residual estimates for the crop specific models. The point estimates on inputs are presented in Appendix Table~\ref{tab:appcrops}.} All models are estimated using the log of yield as the dependent variable and log values of inputs as independent variables.\footnote{Given the prevalence of zero values in the input data, and to a lesser extent in the output data, we use the inverse hyperbolic sine transformation to convert levels to logarithmic values.} Hence, point estimates can be read directly as elasticities.

\subsection{Production Function Results}
The three production functions presented in Table~\ref{tab:reg} rely on the same sample and contain the same set of inputs. To account for heterogeneous input response across crops, we allow all slope and intercept estimates to vary by crop. Model 1 is the classical OLS regression as represented by equation~\eqref{eq:OLS}. This regression model contains fixed effects for seasons but does not account for the multilevel structure of the data. Models 2 and 3 are the MLE and Bayesian regressions and explicitly account for the clustering of observations at the parcel, household, season, village, and temporal levels. The point estimates from the Bayesian regression are based on posterior density estimates derived from iterating from our uninformative priors. Results from these regressions point to a fairly robust set of basic patterns that are repeated with few exceptions. These include (i) positive and significant production relationships between yields and measured inputs (labor, fertilizer, mechanization and pesticides), and (ii) diminishing returns to inputs other than labor. We observe only one instance in which the point estimate for an input is negative and significant (for pesticides in the case of wheat). Returns to scale appear to to be increasing for all crops. The Bayesian estimation of the production function generates point estimates that are broadly similar in sign, magnitude, and significance to those of the OLS and MLE multilevel regressions.

Table~\ref{tab:ICC} reports estimated variance parameters (Panel A), ICCs (Panel B) and variance shares (Panel C) for the multilevel MLE and Bayes regressions. These statistics establish the key findings that inform our insights into the potential role of weather index insurance.\footnote{An expanded set of results from models estimated using different hierarchical structures is provided in Appendix Tables~\ref{tab:appICC} and~\ref{tab:appalt}.} We focus attention on Panel C of Table~\ref{tab:ICC}, which compactly summarizes the data expressed in the upper panels of the table. Reading down the rows of the table allows us to assess the decomposition of variance and, by extension, the relative importance of each level in explaining overall variance in yields. We find for the MLE specification (Model 2 reported in Table~\ref{tab:reg}) that 23 percent of the total variance in yields comes from between-parcel differences, 19 percent is attributed to between-season differences, 15 percent comes from between-village differences, three percent is attributed to differences across time, and 40 percent of the total residual is idiosyncratic noise. In the Bayesian specification (Model 3 reported in Table~\ref{tab:reg}) 24 percent of total variance in yields comes from between-parcel differences, 20 percent is attributed to between-season differences, 15 percent comes from between-village differences, six percent is attributed to differences across time, and 35 percent is idiosyncratic noise. An intuitive interpretation of these results is that much of the differences observed in yields reflects differences between parcels, such as soil quality. Household or farmer capability, relative to other sources of variability, is unimportant in explaining differences in yields. In other words, good farmers cannot make up for bad soil but bad farmers can still prosper if they have good soil. Similar to \cite{Townsend94} and \cite{RosenzweigBinswanger93} we find that idiosyncratic sources of risk play a much larger role in determining observed yields than covariate sources. Unlike \cite{Townsend94} and \cite{RosenzweigBinswanger93}, we are able to quantify these differences. Considering all sources of variance in yields, only 37-41 percent comes from covariate sources while the remaining 59-63 percent comes from idiosyncratic sources. Thinking in terms of insurable weather risk, only 19-20 percent of the variability in crop yield is due to seasonal weather variation. These basic patterns highlight the relatively small importance of between-season yield variance compared with other sources of yield variance.\footnote{Note that the share of variance that remains unexplained is not the appropriate measure of model fit. The Akaike Information Criterion (AIC) and Deviance Information Criterion (DIC) reported in Table~\ref{tab:reg} provides a measure of model fit. In both the AIC and DIC, lower values correspond to better model fit. The DIC is a generalization of the more common Bayesian Information Criterion (BIC) and is used here to facilitate comparison between the ML and Bayesian estimations of the same models.}

Strictly speaking, the non-normality of the error term means that the OLS and MLE regressions are misspecified. Accordingly, we would expect estimates from these models to be biased. However, in our application, only the disturbance terms associated with households and time are highly skewed and non-normal (recall Figure~\ref{fig:zeta}). Furthermore, these two sources of variance make up a very small share of total variance. Thus, any bias that is introduced through model misspecification is mitigated by the small impact of the skewed noise term on the overall estimation. We highlight that this is an artifact of the current application and data. If non-normal variation accounted for more of the total variance we would expect to find greater bias in our OLS and MLE estimates.

Figure~\ref{fig:taui} reports distribution profiles based on the Bayesian regression results for the associated sources of variance measured by Model 3 in Table~\ref{tab:reg}. These histograms are constructed from the posterior distributions estimated using the Gibbs sampler. We can draw several conclusions from this graph. First, several of the variance terms are not normally distributed, which violates the normality assumption in OLS and non-Bayesian multilevel regression. Second, normal confidence interval calculation for the parameters will be inaccurate. Third, the long right-hand side tails in the distributions for households and time, and to a lesser extent season, indicates that these controls by themselves are insufficient to explain the observed yield variability in a classical setting. Said differently, our model relies on indicators for household ability, occurrences of weather, and technical change and does not take into account their severity. The skewed output of our Bayesian analysis hints that while most households are similar in ability there are a few outlying farmers with exceptional ability. Similarly, while weather events account for only 20 percent of variance in yields, severe weather events remain important in considering tools to mitigate risk. Our actuarial analysis in Section~\ref{sec:ins} addresses this limitation in our regression analysis by separately assessing the importance of weather severity in determining the potential benefits of weather index insurance for rural households.

\subsection{Variance Results by Crop Type}
Up to this point, our analysis has accommodated heterogeneity in the input-response curves of the crops under consideration by allowing slopes and intercepts to vary across crops. By pooling these data, however, the variance results reported in Table~\ref{tab:ICC} are implicitly derived under the assumption that the same variance structure of the hierarchical model applies to all crops under consideration and that, by extension, the same weather risk profile applies to each crop. We now relax that assumption and re-estimate the MLE and Bayesian regressions, in each case using five separate, crop-specific sub-samples of the data. Our goal is to allow for a more comprehensive assessment of the variance structure of yields, to gain additional insights into the nature of weather risk exposure among farmers in the sample. The new samples vary in size and coverage, which also permits us to limit the influence of rice in our results, since rice dominates the pooled sample by contributing roughly half of all observations. In three cases (rice, wheat, and maize) we observe data across all 10 seasons. In the case of sorghum and cotton we observe data for only five seasons.

As a complement to the data reported in Table~\ref{tab:ICC}, we present in Table~\ref{tab:crops} a full set of estimated variance parameters, ICCs, and variance shares for the crop-specific versions of our regressions.\footnote{We relegate the discussion of the point estimates for parcel-level inputs to Appendix A and Appendix Table~\ref{tab:appcrops}.} As in the case of the regressions with the pooled samples, the MLE and Bayesian regressions explicitly account for the clustering of observations at the parcel, household, season, village, and temporal levels. The point estimates from the Bayesian regression are based on posterior density estimates derived from iterating from our uninformative priors. As before, we focus attention on Panel C of Table~\ref{tab:crops}. Once again we see a close correspondence between the MLE and Bayesian results. Reading across the table, however, we observe substantial variation in the sources and relative importance of each level in explaining overall variance in crop-specific yields. For example, over 30 percent of variance comes from between-parcel differences for rice and maize, compared with less than two percent for sorghum and cotton. Of particular importance for index insurance, we observe only moderate variation in the contribution of between-season weather differences, which range from 14 percent of variance in the case of rice, to as much as 21 percent in the case of wheat and cotton. We also find more notable differences between the estimated variances from MLE and Bayesian regressions, with implications for the role of non-normality and the inability of the MLE approach to properly account for skewness in several of the disturbance terms. As noted above, to the extent that non-normal errors account for more of the total variance one would expect to find greater bias in the OLS and MLE point estimates vis-\`{a}-vis the Bayes estimates. This is evident when level variance terms tend to be small, resulting in MLE estimates that collapse to zero while Bayesian estimates are small but non-zero.

\section{Actuarial Values of Rainfall-Index Insurance Risk}\label{sec:ins}
The likely importance of extreme and infrequent weather events in yield variability, evidenced by the skewness in the posterior distribution of $\sigma_3^2$, highlights the need for smallholder farmers to have access to affordable risk-management tools. In other words, despite the low share of seasonal weather variation in yield variability, our data still suggest that potential insurance purchasers need to consider weather risk, since extreme weather events, while rare, are nevertheless potentially costly over the long run. This motivates us to ask whether tools for managing this type of risk are adequately accessible. In this section, we complete our analysis of weather risk and agricultural production with a brief quantitative assessment which suggests that rainfall index insurance, as currently marketed in village India, is overpriced.
 
Our parcel-level panel data are not adapted to measuring the severity of weather events. In order to estimate the actuarial cost of these events, we rely on rain-index insurance contracts which ICICI Lombard offers in the region and which form the basis for the analysis conducted by \cite{ColeEtAl13} and \cite{GineEtAl07}. The terms of these contracts are district-dependent, and are the sum of three individual contracts in the three successive phases of the monsoon season. The start and finish dates defining each of the three phases are determined for each village for each year, as a function of the daily evolution of rainfall for the village's corresponding weather station.\footnote{Specifically, Phase I begins on the day when accumulated rainfall since the traditional monsoon start date exceeds $50$mm. In Eastern and Central India, monsoon start date is generally considered 1 June, so Phase I begins for a village when accumulated rainfall since 1 June exceeds $50$mm. In Western India monsoon start date is figured as 1 July. If accumulated rainfall in the first month fails to exceed $50$mm, then the first phase automatically begins on the first of the next month. Phase I and II are each 35 days in length. Phase II begins the day after the close of Phase I. Phase III begins the day after the close of Phase II. In contrast to the first two phases, Phase III lasts for 45 days \citep{WB11}.} In Phases I and II, corresponding to planting and heading, a drought risk exists. In Phase III, corresponding to harvest, risk arises from excess rain. In Phase I and II the contract pays zero Indian Rupees (Rs.) if the cumulative rainfall exceeds a certain ``strike'' value $k$. It then pays a number of Rs. which decreases linearly with the cumulative rainfall in that phase (the value reported in the contract is the slope's absolute value $m$), and if this rainfall amount reaches down to or beyond a certain low ``exit'' number $z$, the contract pays a maximum value $M$, typically $M=1000$ Rs.\footnote{Typically, the payout amount at the exit value $z$, which equals $(k-z)m$, is less than the maximum payout $M$ after exit, though it can sometimes be higher; such an odd structure might be construed as a contract design error, but does occur in commercial practice; we have encountered one instance of this in our analysis of the ICICI Lombard contracts. The probability of the rainfall equaling exactly $z$ millimeters is extremely low, which means that there is usually no ambiguity as to whether to pay the value $(k-z)m$ or the typically larger value $M$. We have identified one important exception to this situation, that in which the exit equals $0$. ICICI Lombard proposes such contracts in relatively arid districts. In that case, the ambiguity would pose a problem, and the company pays the exit max value $M$ when rainfall is $0$. We have taken this into account in our calculations.} The contract is therefore structured so as to have a piecewise linear payout which is typically discontinuous, with a single jump up as rainfall goes from the exit value $z$ to lower values (See Figure 3). In Phase III, the contract is structured in an opposite way: the payout equals zero for cumulative rainfall below the strike value $k$, it equals $M$ (typically $1000$ Rs.) above the exit value $z$, and it grows linearly with positive slope $m$ for cumulative rainfall between $k$ and $z$. The linear increasing or decreasing portions of the contract structure means that its actuarial value will be a bona fide quantitative measure of the severity of drought and/or flood risk. This would not be the case if the contract paid only either zero or a fixed lump sum, since in that case, the actuarial value would only measure the probability of a payout, not drought/flood severity.

We utilize daily rainfall data from 16 of the VDSA villages selected to provide a wide geographic distribution of Indian states with semi-arid, tropical wet-and-dry, and humid subtropical climates from Andhra Pradesh to Bihar for the same years as our regression data, which includes the 2009 drought year. Our method of calculating the actuarial value of the payouts differs from that used in previous studies of the same contracts, which generally take a long view and utilize rainfall data extending as far back as four decades \citep{Parchure02, ClarkeEtAl12a, ColeEtAl13}.\footnote{\cite{ColeEtAl13} provide a straightforward and representative explanation of the standard method of calculation. They utilize 36 years of rainfall data from five villages in Andhra Pradesh and 38 years of rainfall data from three villages in Gujarat to calculate values for each of the three contract phases.} We instead use five years of rainfall data to determine rainfall in each of the three phases in 16 villages. Because of the wide geographic distribution of our data, this averaging encompasses a number of differing weather conditions, which can be considered as a proxy for following a specific village over a long period of time, and ensures that a sufficient number of extreme weather events are included, a particularly useful feature in this region where droughts are sporadic. Additionally, we believe that the shorter time horizon utilized in our calculation more closely approximates the decision horizon for smallholder farmers in India. Given that the average age of the head of household in our data is 49 years, rainfall information from 38 years ago is not likely to figure into insurance purchase decisions.\footnote{Another instance of using wide geographic diversity to study the distribution of rain-indexed insurance in India can be found in \cite{GineEtAl07}, where available commercial premia, strikes,  and exits, were averaged over a number of different districts/weather stations. See Appendix~\ref{sec:appins} for a detailed justification of this approach.}
 
Using our five year time horizon combined with wide geographic diversity, we calculate an estimate of the actuarial value of the payouts for the 3-phase contract by averaging payouts over all $171$ village-year-phase combinations.\footnote{We have five years of rainfall observations for three villages in Andhra Pradesh and two villages in Karnataka. Due to data collection issues the VDSA only reports three years of rainfall data for four villages in each Bihar and Odisha and three villages each in Jharkhand. Given that the contracts under analysis have three phases, this gives us 171 rainfall data points.} To illustrate the effect of how different strike and exit values effect the insurance contract's value, we repeat the calculation for three contracts reported in \cite{GineEtAl07} and \cite{ColeEtAl13}, with low, medium, and high payout structures, and report on the payouts' actuarial values and payout probabilities. The results from this analysis are given in Table~\ref{tab:Ins}.

The three contracts we consider here are representative of products currently marketed to smallholder farmers in the region and were chosen so as to present the most conservative analysis possible.\footnote{We have included in Appendix~\ref{sec:appins} another set of contracts, from \cite{ColeEtAl13} as robustness checks to the values presented in the paper. These contracts have less favorable terms, and would paint a picture of the options available to micro-insurance customers which is arguably too extreme. We have presented our analysis of these less favorably marketed contracts in Appendix Table~\ref{tab:appins} for comparison purposes and for the sake of completeness.} Under the most generous payout structure (``high payout''), the probability of payout is 14 percent. This represents, on average, little more than one payout in one phase over a two-year period. Judging by the examples of contracts reported in \cite{ColeEtAl13}, it is more likely that smallholders would be offered contracts in the medium or low payout ranges, where payout probabilities are 12.3 and 8.8 percent respectively. Farmers who do not look ahead more than three years in planning their activities would presumably find little incentive to purchase insurance where, on average, a 12 percent figure means that they would be as likely as not to see no payout within their planning horizon, and an eight percent figure would br even less favorable. This infrequency of payout on what amounts to income insurance is likely to compound the friction caused by slow diffusion of uptake. For example, studying households in Gujarat, \cite{ColeEtAl14} find that demand for insurance is highly sensitive to whether payouts occurred in the village in the prior year. We suspect that in the villages we consider, only the high payout contract would likely be of interest to farmers, and then only if they perceived its premium as close to fair. Table~\ref{tab:Ins} reports those actuarially fair premia.
 
\cite{ColeEtAl13} report expected payouts as percentages of true premia paid to the ICICI Lombard insurance company for two villages in Andhra Pradesh in 2006. These numbers are equivalent to saying that the premium is computed by multiplying the actuarially fair premium by a loading factor, $1 + \lambda$, through the formula:

\begin{equation}
1 + \lambda = \frac{\textrm{paid price}}{\textrm{actuarially fair price}}.
\end{equation}

\noindent Thus $\lambda$ is a proportional transaction cost which, in principle, reflects the cost to the company of doing business. The loading factors' transaction cost parameters computed from reported premia ($1.47$, $2.32$, $3.86$), are very large compared with other weather insurance markets: a typical value of $\lambda$ in the US and other countries is between $0.1$ and $0.2$. For instance, the Poverty Global Practice Group at the World Bank reports values in the US, Burkina Faso, and Senegal of $\lambda =0.10$ \citep{Nicola15}. While it may not be possible for insurers to offer contracts with such low values of $\lambda$ in India, our actuarially fair value for the high-payout contract, $191$ Rs., can be compared directly to the price quoted by ICICI Lombard in \cite{ColeEtAl13} for the same contract, which is $280$ Rs. In this case, which is the most favorable comparison we can draw with commercial contract prices, the high-payout contract would thus represent a loading factor of $1.47$, i.e. a value of $\lambda = 0.47$, a substantially higher value than those observed in other regions, but still much lower than those inferred from the two villages reported in \cite{ColeEtAl13}. Such excessive loading factors, and not apparent failure of households to recognize the value of insurance, appear to be a key element in preventing farmers in village India from accessing insurance as a risk-management resource and further explains the low uptake rate of weather index insurance.

\section{Conclusion}\label{sec:conclusion}
Despite long standing evidence that rural households are unable to fully insure covariate risk, few studies have attempted to measure just how large a role covariate events, such as weather, play in agricultural yields. We address this research gap using agricultural production data covering 11,942 parcel level observations from India. Using a multilevel/hierarchical regression framework, we estimate the different sources of yield variance. This approach controls for inputs at the parcel-level and also isolate the amount of yield variability attributed to parcel-level effects, household-level effects, seasonal weather effects, village-level effects, and time effects. Adopting Bayesian estimation techniques allows us to account for the highly skewed distribution of several of the disturbance terms. Overall, we find that variability in weather makes up only a small share of the total variance in crop yields. This suggests that basis risk on the production side (i.e., low correlation between the weather index and yield loss) is substantial. The majority of variation in yields does not come from seasonal variability, which is our proxy for rainfall or weather variability. Rather, the majority of variation in yields comes from differences in parcels and from the random disturbance term which captures idiosyncratic events which would be, by definition, not covered by index insurance. Given the long history of evidence that farmers are effective in using self-insurance and mitigation measures to protect against idiosyncratic risk and given the small role covariate seasonal risk plays in crop variability, farmers may rationally prefer to forgo weather index insurance at any positive price to focus their risk management choices on minimizing other sources of risk.

Although there are many potential impediments to insurance uptake, this research provides evidence for an obvious but until now unsupported explanation: insurance contracts are overpriced from the perspective of the farming household. Product design and ratemaking of index insurance in India has generally taken a long view, utilizing rainfall data extending as far back as four decades to calculate actuarial rates. From the perspective of the insurance company, such a long time horizon may support the company's longevity in managing risk. However, it may not easily align with the planning horizons of rural households. Motivated by our econometric results, we take a shorter-term perspective and calculate the actuarial value of payouts using rainfall data from five seasons, including a drought season, and a large number of geographically dispersed locations. We find that loading factors on these contracts are excessive and that this pricing problem undermines the incentives for smallholder farmers to purchase insurance. Our combines results suggest that micro-insurance markets could hold limited promise for improving household risk management, and then only to the extent that future products can adapt to the realities facing potential purchasers.

\newpage
\singlespace
\bibliographystyle{chicago}
\bibliography{MLMref}

\newpage

\begin{table}[!htbp] \centering 
\caption{Descriptive Statistics by Crop} \label{tab:cropdisc}
\setlength{\linewidth}{.1cm}\newcommand{\contents}
{\begin{tabular}{l cccccc}
\\[-1.8ex]\hline 
\hline \\[-1.8ex] 
&        Rice&     Sorghum&       Wheat&       Maize&      Cotton&       Total\\
\midrule
yield (kg/ha)       &        3,356&        2,368&        1,537&        2,596&        1,383&        2,658\\
                    &      (3,252)&      (1,429)&      (4,038)&      (2,397)&       (870.4)&      (2,956)\\
labor (hr/ha)       &       880.1&       301.6&       760.5&       831.7&      1156.6&       749.8\\
                    &     (825.8)&     (265.9)&    (2,777)&     (769.6)&     (644.4)&    (1,239)\\
fertilizer (kg/ha)  &       215.0&       289.3&       174.6&       252.3&       333.9&       239.2\\
                    &     (213.6)&     (191.9)&     (724.0)&     (224.6)&     (210.0)&     (333.4)\\
mechanization (Rs/ha)&      1,362&      2,891&       805.6&       878.7&      2,927&      1,716\\
                    &    (2,781)&    (4,887)&    (1,289)&    (1,755)&    (3,778)&    (3,369)\\
pesticide (Rs/ha)   &        61.7&       124.5&        57.2&        59.8&      2,066&       235.0\\
                    &     (266.7)&     (566.3)&     (345.9)&     (319.0)&    (2,748)&    (1,012)\\
parcel area (ha)    &       0.355&       0.597&       0.674&       0.317&       0.904&       0.494\\
                    &     (0.501)&     (0.833)&     (0.733)&     (0.334)&     (0.662)&     (0.652)\\
rainfall (mm)       &       539.1&       395.0&       286.5&       456.3&       426.3&       468.4\\
                    &     (253.3)&     (166.6)&      (83.1)&     (192.2)&     (178.1)&     (234.8)\\\midrule 
number of observations& \multicolumn{1}{c}{5,572} & \multicolumn{1}{c}{2,720} & \multicolumn{1}{c}{1,625} & \multicolumn{1}{c}{1,073} & \multicolumn{1}{c}{952} & \multicolumn{1}{c}{11,942} \\ 
number of parcels			& \multicolumn{1}{c}{2,678} & \multicolumn{1}{c}{1,636} & \multicolumn{1}{c}{1,151} & \multicolumn{1}{c}{549} & \multicolumn{1}{c}{600} & \multicolumn{1}{c}{5,100} \\ 
number of households		& \multicolumn{1}{c}{612} & \multicolumn{1}{c}{491} & \multicolumn{1}{c}{383} & \multicolumn{1}{c}{268} & \multicolumn{1}{c}{254} & \multicolumn{1}{c}{1,079} \\ 
number of seasons		& \multicolumn{1}{c}{106} & \multicolumn{1}{c}{81} & \multicolumn{1}{c}{69} & \multicolumn{1}{c}{86} & \multicolumn{1}{c}{51} & \multicolumn{1}{c}{240} \\ 
number of villages		& \multicolumn{1}{c}{21} & \multicolumn{1}{c}{20} & \multicolumn{1}{c}{20} & \multicolumn{1}{c}{20} & \multicolumn{1}{c}{12} & \multicolumn{1}{c}{30} \\ 
number of time periods		& \multicolumn{1}{c}{10} & \multicolumn{1}{c}{5} & \multicolumn{1}{c}{10} & \multicolumn{1}{c}{10} & \multicolumn{1}{c}{5} & \multicolumn{1}{c}{10} \\ 
\\[-1.8ex]\hline 
\hline \\[-1.8ex] 
\multicolumn{7}{p{\linewidth}}{\footnotesize \textit{Note}: Table displays means of data by crop with standard deviations in parenthesis. All monetary values are in real $2010$ Indian Rupees (Rs).}
\end{tabular}}
\setbox0=\hbox{\contents}
\setlength{\linewidth}{\wd0-2\tabcolsep-.25em}
\contents
\end{table}

\begin{table}[!htbp] \centering 
	\caption{Results of Production Function Regressions}   \label{tab:reg}  
	\scalebox{0.66} {\setlength{\linewidth}{.1cm}\newcommand{\contents}
		{\begin{tabular}{@{\extracolsep{5pt}}lD{.}{.}{-3} D{.}{.}{-3} D{.}{.}{-3} } 
				\\[-1.8ex]\hline 
				\hline \\[-1.8ex] 
				\\[-1.8ex] & \multicolumn{1}{c}{\textit{OLS}}	& \multicolumn{1}{c}{\textit{MLE}}  & \multicolumn{1}{c}{\textit{Bayes}}	\\ 
				\\[-1.8ex] & \multicolumn{1}{c}{(1)} 			& \multicolumn{1}{c}{(2)} 			& \multicolumn{1}{c}{(3)}   			\\ 
				\midrule
				\multicolumn{4}{l}{rice} \\
				\cline{1-1}
				log labor 			& 1.635^{***} & 1.531^{***} & 1.127^{***} \\ 
									& (0.043) & (0.043) & (0.032) \\ 
				log fertilizer		& 0.598^{***} & 0.512^{***} & 0.542^{***} \\ 
									& (0.019) & (0.019) & (0.018) \\ 
				log mechanization	& 0.206^{***} & 0.176^{***} & 0.147^{***} \\ 
									& (0.010) & (0.009) & (0.009) \\ 
				log pesticides		& 0.123^{***} & 0.093^{***} & 0.088^{***} \\ 
									& (0.011) & (0.010) & (0.010) \\ 
				
				\multicolumn{4}{l}{sorghum} \\
				\cline{1-1}
				log labor 			& 0.585^{***} & 0.559^{***} & 0.987^{***} \\ 
									& (0.063) & (0.060) & (0.042) \\ 
				log fertilizer		& 0.087^{**} & 0.108^{***} & 0.214^{***} \\ 
									& (0.042) & (0.040) & (0.038) \\ 
				log mechanization 	& 0.321^{***} & 0.342^{***} & 0.491^{***} \\ 
									& (0.030) & (0.029) & (0.026) \\ 
				log pesticides		& 0.009 & 0.012 & 0.009 \\ 
									& (0.018) & (0.016) & (0.016) \\ 

				\multicolumn{4}{l}{wheat} \\
				\cline{1-1}
				log labor 			& 0.993^{***} & 0.988^{***} & 1.217^{***} \\ 
									& (0.054) & (0.053) & (0.036) \\ 
				log fertilizer		& 0.110^{***} & 0.108^{***} & 0.073^{***} \\ 
									& (0.021) & (0.020) & (0.036) \\ 
				log mechanization 	& 0.386^{***} & 0.350^{***} & 0.366^{***} \\ 
									& (0.019) & (0.018) & (0.036) \\ 
				log pesticides		& -0.059^{**} & -0.069^{***} & -0.100^{***} \\ 
									& (0.026) & (0.025) & (0.025) \\ 

				\multicolumn{4}{l}{maize} \\
				\cline{1-1}
				log labor 			& 1.126^{***} & 1.174^{***} & 1.584^{***} \\ 
									& (0.100) & (0.097) & (0.043) \\ 
				log fertilizer		& 0.048 & 0.016 & -0.005 \\ 
									& (0.041) & (0.039) & (0.040) \\ 
				log mechanization 	& 0.111^{***} & 0.096^{***} & 0.109^{***} \\ 
									& (0.018) & (0.017) & (0.017) \\ 
				log pesticides		& 0.086^{***} & 0.094^{***} & 0.098^{***} \\ 
									& (0.033) & (0.030) & (0.029) \\ 

				\multicolumn{4}{l}{cotton} \\
				\cline{1-1}
				log labor 			& 1.047^{***} & 1.064^{***} & 1.291^{***} \\ 
									& (0.129) & (0.124) & (0.057) \\ 
				log fertilizer		& 0.220^{***} & 0.230^{***} & 0.230^{***} \\ 
									& (0.057) & (0.053) & (0.053) \\ 
				log mechanization	 & 0.072^{**} & 0.033 & 0.036 \\ 
									& (0.031) & (0.031) & (0.029) \\ 
				log pesticides		& 0.128^{***} & 0.115^{***} & 0.095^{***} \\ 
									& (0.029) & (0.028) & (0.027) \\ 
				\hline \\[-1.8ex] 
				observations 		& \multicolumn{1}{c}{11,942}	& \multicolumn{1}{c}{11,942}  	& \multicolumn{1}{c}{11,942}	\\ 
				R$^{2}$ 			& \multicolumn{1}{c}{0.964} 	& 								& 								\\ 
				Log Likelihood 		&  								& \multicolumn{1}{c}{-22,268}	& 								\\ 
				Akaike Inf. Crit. 	&  								& \multicolumn{1}{c}{44,598} 	& 								\\ 
				Deviance Inf. Crit.	&  								& \multicolumn{1}{c}{44,828}   	& \multicolumn{1}{c}{42,578}	\\ 
				\hline 
				\hline \\[-1.8ex] 
				\multicolumn{4}{p{\linewidth}}{\footnotesize \textit{Note}: Dependent variable is log of yield. All specifications include statistically significant crop-specific intercept terms as controls. Column (1) is a classical OLS regression that includes season fixed effects but does not account for the multilevel structure of the data. Column (2) is the maximum likelihood estimate of the multilevel model and contains covariates and data clustered at parcel, household, season, village, and time. Column (3) is the Bayesian estimation of the multilevel model. Bayesian calculations use a burn-in period of 5,000 iterations and an additional 5,000 iterations to ensure convergence. Standard errors are reported in parentheses ($^{*}$p$<$0.1; $^{**}$p$<$0.05; $^{***}$p$<$0.01).} \\ 
			\end{tabular}}
			\setbox0=\hbox{\contents}
			\setlength{\linewidth}{\wd0-2\tabcolsep-.25em}
			\contents}
\end{table}

\begin{table}[!htbp] \centering 
\caption{Estimated Variance, ICCs, and Variance Shares from Multilevel Regressions} \label{tab:ICC} 
\setlength{\linewidth}{.1cm}\newcommand{\contents}
{\begin{tabular}{@{\extracolsep{5pt}}lD{.}{.}{-3} D{.}{.}{-3}   } 
	\\[-1.8ex]\hline 
	\hline \\[-1.8ex] 
	\\[-1.8ex] 	&  \multicolumn{1}{c}{\textit{MLE}} &  \multicolumn{1}{c}{\textit{Bayes}} \\ 
	\midrule
	& &  \\
	\multicolumn{3}{c}{Panel A: \textit{Variance Parameter Estimates}}	\\  
	& &  \\
	parcel $(\sigma_{1}^2)$			& 0.933	& 1.098	 	\\ 
	household $(\sigma_{2}^2)$		& 0.002	& 0.004	 	\\ 
	season $(\sigma_{3}^2)$		& 0.790	& 0.903		\\ 
	village $(\sigma_{4}^2)$		& 0.623	& 0.682		\\  
	time $(\sigma_{5}^2)$			& 0.126	& 0.261		\\ 
	idiosyncratic $(\sigma^2)$		& 1.623	& 1.613		\\ 
	\midrule
	& &  \\
	\multicolumn{3}{c}{Panel B: \textit{Intraclass Correlation Coefficients}}	\\ 
	& &  \\	
	parcel			& 0.228	& 0.241		\\	
	household		& 0.228	& 0.242		\\
	season			& 0.421	& 0.440		\\
	village			& 0.573	& 0.589		\\
	time			& 0.604	& 0.646		\\
	\midrule
	& &  \\
	\multicolumn{3}{c}{Panel C: \textit{Shares of Variance From Each Level}}	\\ 
	& &  \\
	parcel			& \multicolumn{1}{c}{23\%}	& \multicolumn{1}{c}{24\%}	\\
	household		& \multicolumn{1}{c}{00\%}	& \multicolumn{1}{c}{00\%}	\\
	season			& \multicolumn{1}{c}{19\%}	& \multicolumn{1}{c}{20\%}	\\
	village			& \multicolumn{1}{c}{15\%}	& \multicolumn{1}{c}{15\%}	\\
	time			& \multicolumn{1}{c}{03\%}	& \multicolumn{1}{c}{06\%}	\\
	idiosyncratic	& \multicolumn{1}{c}{40\%}	& \multicolumn{1}{c}{35\%}	\\
	\midrule
	observations	& \multicolumn{2}{c}{11,942}	\\
	\hline 
	\hline \\[-1.8ex] 
	\multicolumn{3}{p{\linewidth}}{\footnotesize \textit{Note}: In Panel A, estimates of the variance parameters on each level's residuals come from estimation of models reported in Columns (2) and (3) in Table~\ref{tab:reg}. Variances $\sigma_{1}^2$, $\sigma_{2}^2$, $\sigma_{3}^2$, $\sigma_{4}^2$, $\sigma_{5}^2$ represent the variance in crop yield that comes from the corresponding level. The final variance parameter $(\sigma^2)$ corresponds to the idiosyncratic or unexplained portion of the model. Intraclass correlation coefficients in Panel B are calculated using the formulas in equations~\eqref{eq:icc1}-\eqref{eq:icc5}. Panel C decomposes the ICC into percent of variance accorded to each level. For comparison, Appendix Tables~\ref{tab:appICC} and~\ref{tab:appalt} reports alternative specifications of the models presented here.} \\ 
			\end{tabular}}
			\setbox0=\hbox{\contents}
			\setlength{\linewidth}{\wd0-2\tabcolsep-.25em}
			\contents
\end{table}

\begin{landscape}
\begin{table}[!htbp] \centering 
	\caption{Estimated Variance, ICCs, and Variance Shares by Crop} \label{tab:crops} 
	\scalebox{.8}
	{\setlength{\linewidth}{.1cm}\newcommand{\contents}
	{\begin{tabular}{@{\extracolsep{5pt}}lD{.}{.}{-3} D{.}{.}{-3}  D{.}{.}{-3} D{.}{.}{-3} D{.}{.}{-1} D{.}{.}{-3}  D{.}{.}{-3} D{.}{.}{-3} D{.}{.}{-1}  D{.}{.}{-3} D{.}{.}{-3} D{.}{.}{-3}} 
		\\[-1.8ex]\hline 
		\hline \\[-1.8ex] 
		& \multicolumn{2}{c}{Rice}		& \multicolumn{2}{c}{Sorghum}	& \multicolumn{2}{c}{Wheat}		& \multicolumn{2}{c}{Maize}		& \multicolumn{2}{c}{Cotton}	\\ 
		& \multicolumn{1}{c}{\textit{MLE}} &	\multicolumn{1}{c}{\textit{Bayes}} & \multicolumn{1}{c}{\textit{MLE}} &	\multicolumn{1}{c}{\textit{Bayes}} & \multicolumn{1}{c}{\textit{MLE}} &	\multicolumn{1}{c}{\textit{Bayes}} & \multicolumn{1}{c}{\textit{MLE}} &	\multicolumn{1}{c}{\textit{Bayes}} & \multicolumn{1}{c}{\textit{MLE}} &	\multicolumn{1}{c}{\textit{Bayes}} \\
		& \multicolumn{1}{c}{(1)} &	\multicolumn{1}{c}{(2)} & \multicolumn{1}{c}{(3)} &	\multicolumn{1}{c}{(4)} & \multicolumn{1}{c}{(5)} &	\multicolumn{1}{c}{(6)} & \multicolumn{1}{c}{(7)} &	\multicolumn{1}{c}{(8)} & \multicolumn{1}{c}{(9)} &	\multicolumn{1}{c}{(10)} \\
		\midrule
		& & & & & & & & & & \\
		\multicolumn{11}{c}{Panel A: \textit{Variance Parameter Estimates}}	\\ 
		& & & & & & & & & & \\
		parcel $(\sigma_{1}^2)$			& 1.858	& 1.852	& 0.000 & 0.002	& 1.092 & 1.059	& 0.564 & 0.557	& 0.000	& 0.018	\\
		household $(\sigma_{2}^2)$		& 0.000	& 0.003	& 0.048 & 0.046	& 0.046 & 0.034	& 0.000 & 0.007	& 0.013	& 0.010	\\
		season $(\sigma_{3}^2)$		& 0.832	& 0.865	& 0.166 & 0.174	& 2.323 & 2.573	& 0.234 & 0.266	& 0.367	& 0.401	\\
		village $(\sigma_{4}^2)$		& 1.098	& 1.266	& 0.133 & 0.156	& 5.527 & 6.411	& 0.185 & 0.177	& 0.000	& 0.041	\\
		time $(\sigma_{5}^2)$			& 0.203	& 0.298	& 0.000 & 0.011	& 0.612 & 0.694	& 0.000 & 0.013	& 0.521	& 0.689	\\
		idiosyncratic $(\sigma^2)$		& 1.918	& 1.918	& 0.686 & 0.684	& 1.598 & 1.626	& 0.517 & 0.518	& 0.785	& 0.766	\\
		\midrule
		& & & & & & & & & & \\
		\multicolumn{11}{c}{Panel B: \textit{Intraclass Correlation Coefficients}}	\\ 
		& & & & & & & & & & \\
		parcel							& 0.314	& 0.299	& 0.000 & 0.002	& 0.098 & 0.085	& 0.376 & 0.362	& 0.000	& 0.009	\\
		household						& 0.314	& 0.299	& 0.046 & 0.045	& 0.102 & 0.088	& 0.376 & 0.366	& 0.008	& 0.015	\\
		season							& 0.455	& 0.439	& 0.207 & 0.207	& 0.309 & 0.296	& 0.532 & 0.539	& 0.225	& 0.223	\\
		village							& 0.641	& 0.643	& 0.336 & 0.352	& 0.803 & 0.813	& 0.655 & 0.654	& 0.225	& 0.244	\\			
		time							& 0.675	& 0.691	& 0.336 & 0.363	& 0.857 & 0.869	& 0.655 & 0.663	& 0.534	& 0.602	\\
		\midrule
		& & & & & & & & & & \\
		\multicolumn{11}{c}{Panel C: \textit{Shares of Variance From Each Level}}	\\ 
		& & & & & & & & & & \\
		parcel			& \multicolumn{1}{c}{31\%}		& \multicolumn{1}{c}{30\%}		& \multicolumn{1}{c}{00\%}		& \multicolumn{1}{c}{00\%}		& \multicolumn{1}{c}{10\%} & \multicolumn{1}{c}{09\%}		& \multicolumn{1}{c}{38\%}		& \multicolumn{1}{c}{36\%}		& \multicolumn{1}{c}{00\%}		& \multicolumn{1}{c}{01\%}		\\
		household		& \multicolumn{1}{c}{00\%}		& \multicolumn{1}{c}{00\%}		& \multicolumn{1}{c}{05\%}		& \multicolumn{1}{c}{04\%}		& \multicolumn{1}{c}{00\%} & \multicolumn{1}{c}{00\%}		& \multicolumn{1}{c}{00\%}		& \multicolumn{1}{c}{00\%}		& \multicolumn{1}{c}{01\%}		& \multicolumn{1}{c}{01\%}		\\
		season			& \multicolumn{1}{c}{14\%}		& \multicolumn{1}{c}{14\%}		& \multicolumn{1}{c}{16\%}		& \multicolumn{1}{c}{16\%}		& \multicolumn{1}{c}{21\%} & \multicolumn{1}{c}{21\%}		& \multicolumn{1}{c}{16\%}		& \multicolumn{1}{c}{17\%}		& \multicolumn{1}{c}{22\%}		& \multicolumn{1}{c}{21\%}		\\
		village			& \multicolumn{1}{c}{19\%}		& \multicolumn{1}{c}{20\%}		& \multicolumn{1}{c}{13\%}		& \multicolumn{1}{c}{15\%}		& \multicolumn{1}{c}{49\%} & \multicolumn{1}{c}{52\%}		& \multicolumn{1}{c}{12\%}		& \multicolumn{1}{c}{12\%}		& \multicolumn{1}{c}{00\%}		& \multicolumn{1}{c}{02\%}		\\
		time			& \multicolumn{1}{c}{03\%}		& \multicolumn{1}{c}{05\%}		& \multicolumn{1}{c}{00\%}		& \multicolumn{1}{c}{01\%}		& \multicolumn{1}{c}{05\%} & \multicolumn{1}{c}{06\%}		& \multicolumn{1}{c}{00\%}		& \multicolumn{1}{c}{01\%}		& \multicolumn{1}{c}{31\%}		& \multicolumn{1}{c}{36\%}		\\
		idiosyncratic	& \multicolumn{1}{c}{32\%}		& \multicolumn{1}{c}{31\%}		& \multicolumn{1}{c}{66\%}		& \multicolumn{1}{c}{64\%}		& \multicolumn{1}{c}{14\%} & \multicolumn{1}{c}{13\%}		& \multicolumn{1}{c}{34\%}		& \multicolumn{1}{c}{34\%}		& \multicolumn{1}{c}{47\%}		& \multicolumn{1}{c}{40\%}	\\
		\midrule
		observations	& \multicolumn{2}{c}{5,572}	& \multicolumn{2}{c}{2,720}		& \multicolumn{2}{c}{1,625}		& \multicolumn{2}{c}{1,073}		& \multicolumn{2}{c}{952}		\\
		\hline 
		\hline \\[-1.8ex] 
		\multicolumn{11}{p{\linewidth}}{\footnotesize \textit{Note}: In Panel A, estimates of the variance parameters on each level's residuals come from estimation of models in corresponding columns in Appendix Table~\ref{tab:appcrops}. Variances $\sigma_{1}^2$, $\sigma_{2}^2$, $\sigma_{3}^2$, $\sigma_{4}^2$, $\sigma_{5}^2$ represent the variance in crop yield that comes from the corresponding level. The final variance parameter $(\sigma^2)$ corresponds to the idiosyncratic or unexplained portion of the model. Intraclass correlation coefficients in Panel B are calculated using the formulas in equations~\eqref{eq:icc1}-\eqref{eq:icc5}. Panel C decomposes the ICC into percent of variance accorded to each level. Bayesian calculations use a burn-in period of 5,000 iterations and an additional 5,000 iterations to ensure convergence.} \\ 
	\end{tabular}}
	\setbox0=\hbox{\contents}
	\setlength{\linewidth}{\wd0-2\tabcolsep-.25em}
	\contents}
\end{table}
\end{landscape}

\begin{table}[!htbp] \centering 
  \caption{Actuarial Values and Payout Probabilities} \label{tab:Ins} 
	\scalebox{.8}
	{\setlength{\linewidth}{.1cm}\newcommand{\contents}
	{\begin{tabular}{@{\extracolsep{1pt}}lD{.}{.}{-3}D{.}{.}{-3} D{.}{.}{-3} D{.}{.}{-3} D{.}{.}{-3} D{.}{.}{-3} D{.}{.}{-3}}
\\[-1.8ex]\hline 
\hline \\[-1.8ex]
  & \multicolumn{3}{c}{\textit{Contract Structure}} & \multicolumn{4}{c}{\textit{Summary Statistics}}  \\ 
  &  &  &  & \multicolumn{1}{c}{Actuarially Fair} & \multicolumn{1}{c}{Probability} & \multicolumn{1}{c}{Loading} & \multicolumn{1}{c}{Years until} \\ 
& \multicolumn{1}{c}{Strike} & \multicolumn{1}{c}{Exit} & \multicolumn{1}{c}{Max Payout} & \multicolumn{1}{c}{Premium} & \multicolumn{1}{c}{of Payout} & \multicolumn{1}{c}{Factor} & \multicolumn{1}{c}{Payout} \\ 
\midrule
& & & & & & & \\
\multicolumn{8}{c}{Panel A}	\\ 
\multicolumn{8}{c}{\textit{High-Payout Contract Structure}}	\\ 
& & & & & & & \\
\multicolumn{1}{l}{Phase I}		& 70	& 10	& 1000	& \multirow{3}{*}{190.9}	& \multirow{3}{*}{14.0\%}  & \multirow{3}{*}{1.47}	& \multirow{3}{*}{2.38}  \\ 
\multicolumn{1}{l}{Phase II}	& 80	& 10	& 1000	&	&  &  &  \\ 
\multicolumn{1}{l}{Phase III} 	& 375	& 450	& 1000	&  	&  &  &  \\ 
\midrule
& & & & & & & \\
\multicolumn{8}{c}{Panel B}	\\ 
\multicolumn{8}{c}{\textit{Medium-Payout Contract Structure}}	\\ 
& & & & & & & \\
\multicolumn{1}{l}{Phase I}		& 78	& 15		& 1000	& \multirow{3}{*}{129.2}	& \multirow{3}{*}{12.3\%}  & \multirow{3}{*}{2.32}	& \multirow{3}{*}{2.71}  \\ 
\multicolumn{1}{l}{Phase II}	& 72	& 12		& 1000	&  &  &  &  \\ 
\multicolumn{1}{l}{Phase III}	& 499	& 580		& 1000	&  &  &  &  \\ 
\midrule
& & & & & & & \\
\multicolumn{8}{c}{Panel C}	\\ 
\multicolumn{8}{c}{\textit{Low-Payout Contract Structure}}	\\ 
& & & & & & & \\
\multicolumn{1}{l}{Phase I}		& 50	& 5		& 1000	& \multirow{3}{*}{69.6}	& \multirow{3}{*}{8.77\%} & \multirow{3}{*}{3.86}	& \multirow{3}{*}{3.80}   \\ 
\multicolumn{1}{l}{Phase II}	& 60	& 5		& 1000	&  &  &  &  \\ 
\multicolumn{1}{l}{Phase III}	& 560	& 670 	& 1000 	&  &  &  &  \\ 
 \\[-1.8ex]\hline 
\hline \\[-1.8ex] 
\multicolumn{8}{p{\linewidth}}{\footnotesize \textit{Note}: High- and low-payout contracts come from \cite{ColeEtAl13} while the medium-payout contract comes from \cite{GineEtAl07}. While each contract from \cite{ColeEtAl13} is designed for a specific village/weather station, we follow \cite{GineEtAl07} in utilizing a large representative set of rainfall data to calculate actuarially fair premia and the probability of payout. Actuarially fair premia are calculated using standard actuarial principles. Payout probability is the average occurrence of a payout. Loading factors are calculated as $1 + \lambda = \frac{\textrm{paid price}}{\textrm{actuarially fair price}}$. Years until payout are calculated by dividing the inverse of the probability of payout by three, which assumes payout events occur as a Poisson process. For comparison, Appendix Table~\ref{tab:appins} reports the same calculations for the remaining three contracts reported in \cite{ColeEtAl13}.} \\ 
\end{tabular}}
\setbox0=\hbox{\contents}
\setlength{\linewidth}{\wd0-2\tabcolsep-.25em}
\contents}
\end{table}

\newpage
\begin{landscape}
\begin{figure}[!htbp] 
\includegraphics[width=.9\linewidth,keepaspectratio]{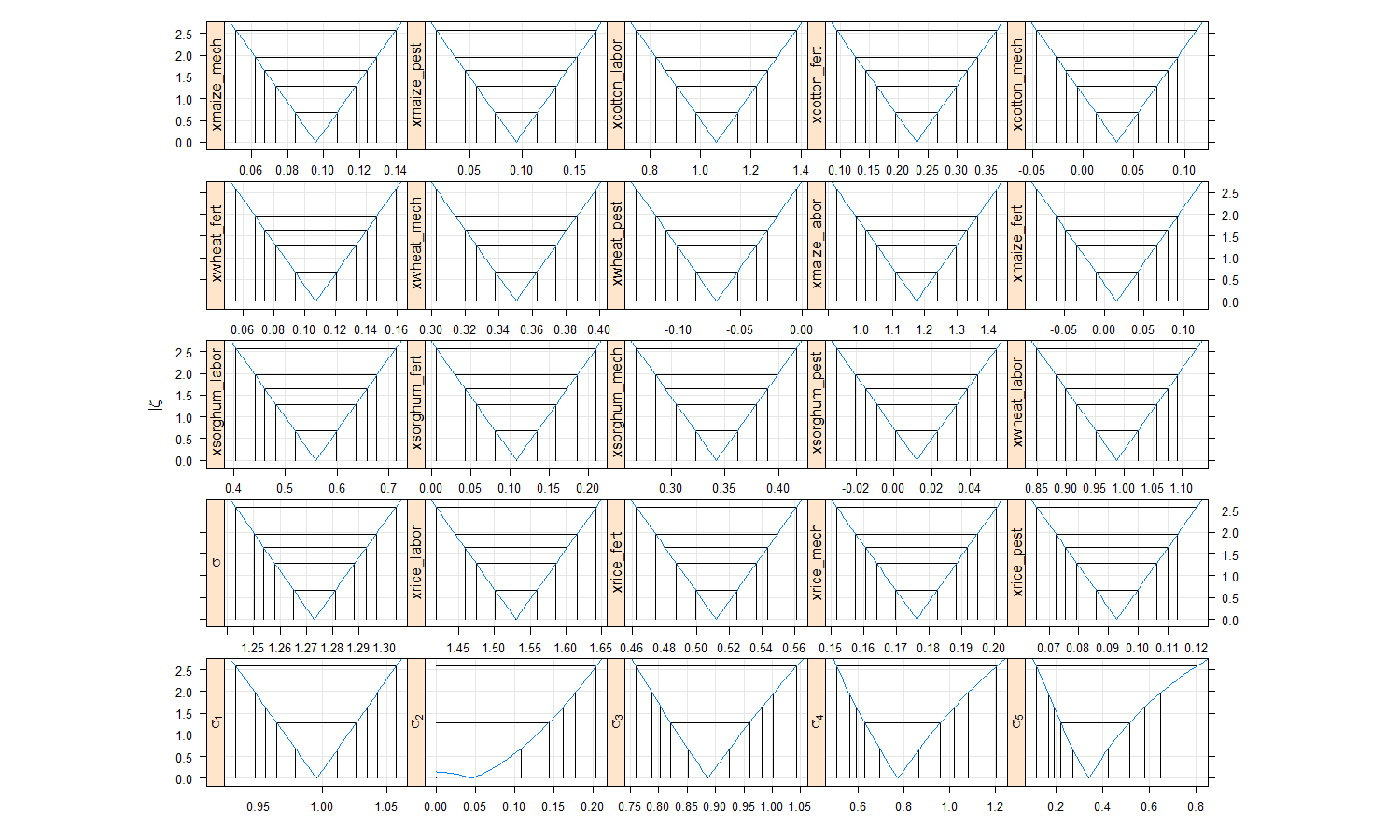}
\caption{Zeta Profile Plots for Multilevel Model} \label{fig:zeta}
\footnotesize \textit{Note}: The figure presents the zeta profile plots, representing the underlying distribution of the parameters resulting from the model fit to equations~\eqref{eq:level0}-\eqref{eq:level5}. Plots are calculated by first estimating the model. Then, holding a single parameter fixed, the other parameters are varied to assess the fit of each new iteration compared to the globally optimal fit. The comparison statistic is the likelihood ratio test. Finally, a signed square root transformation is applied to the LR statistic. The plots are the absolute value of the resulting function, $| \zeta |$, in comparison to the estimated parameter values. Parameters with underlying normal distributions have straight line zeta profile plots. The vertical lines delimit $50\%$, $80\%$, $90\%$, $95\%$, and $99\%$ confidence intervals. For zeta profile plot lines that are not straight, the normal distribution is a poor approximation of the underlying distribution.
\end{figure}

\newpage
\begin{figure}[!htbp]
\centering

\includegraphics[width=.45\linewidth,keepaspectratio]{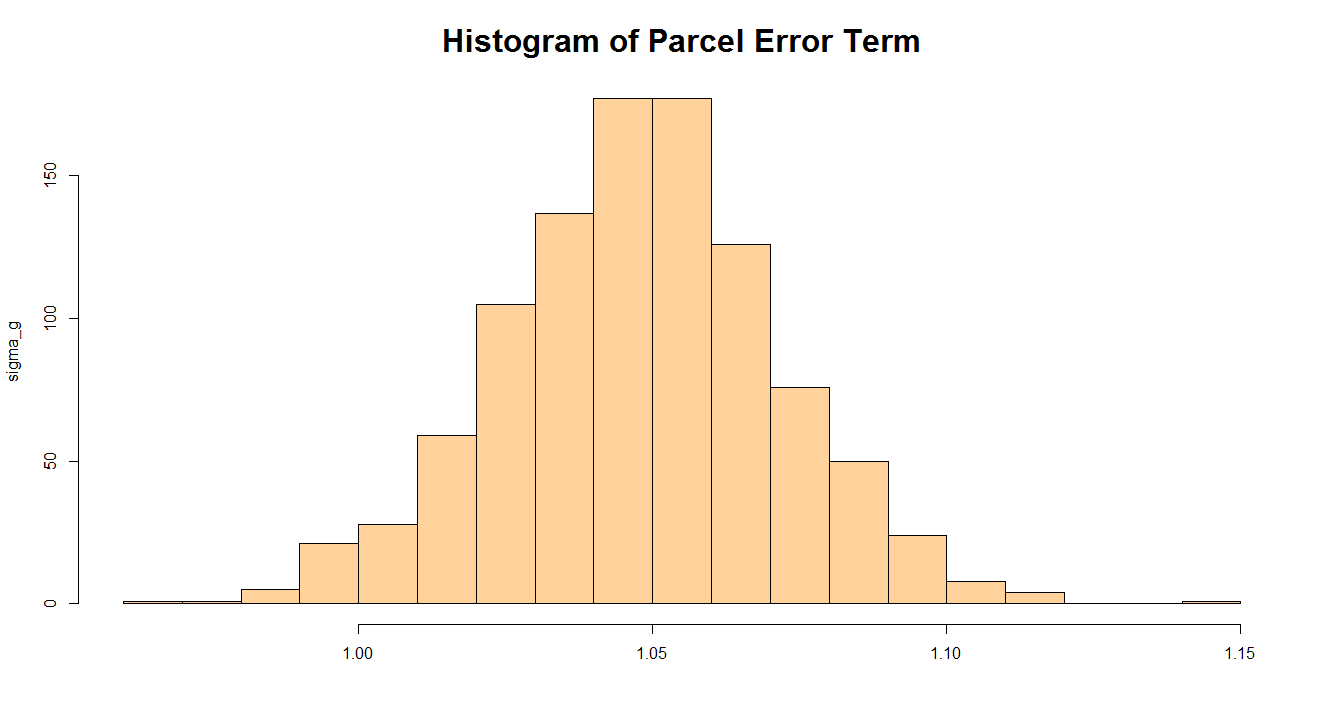}
\includegraphics[width=.45\linewidth,keepaspectratio]{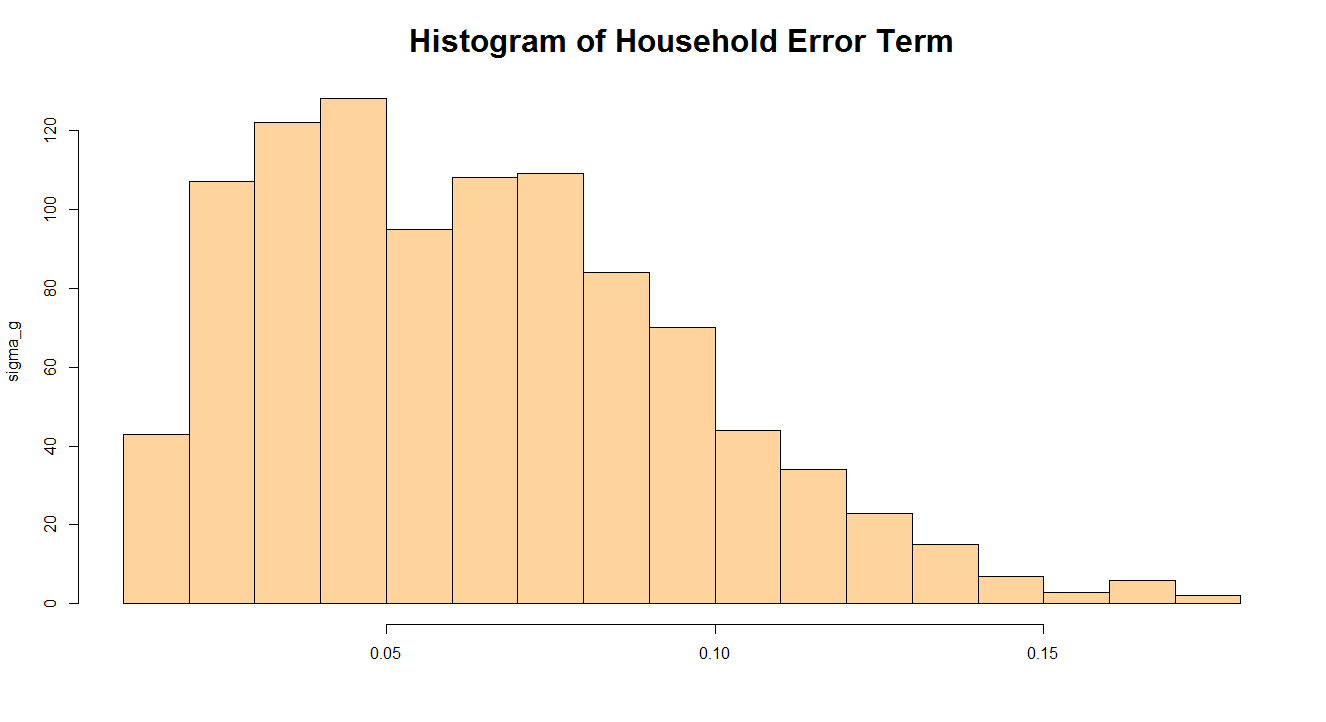}
\includegraphics[width=.45\linewidth,keepaspectratio]{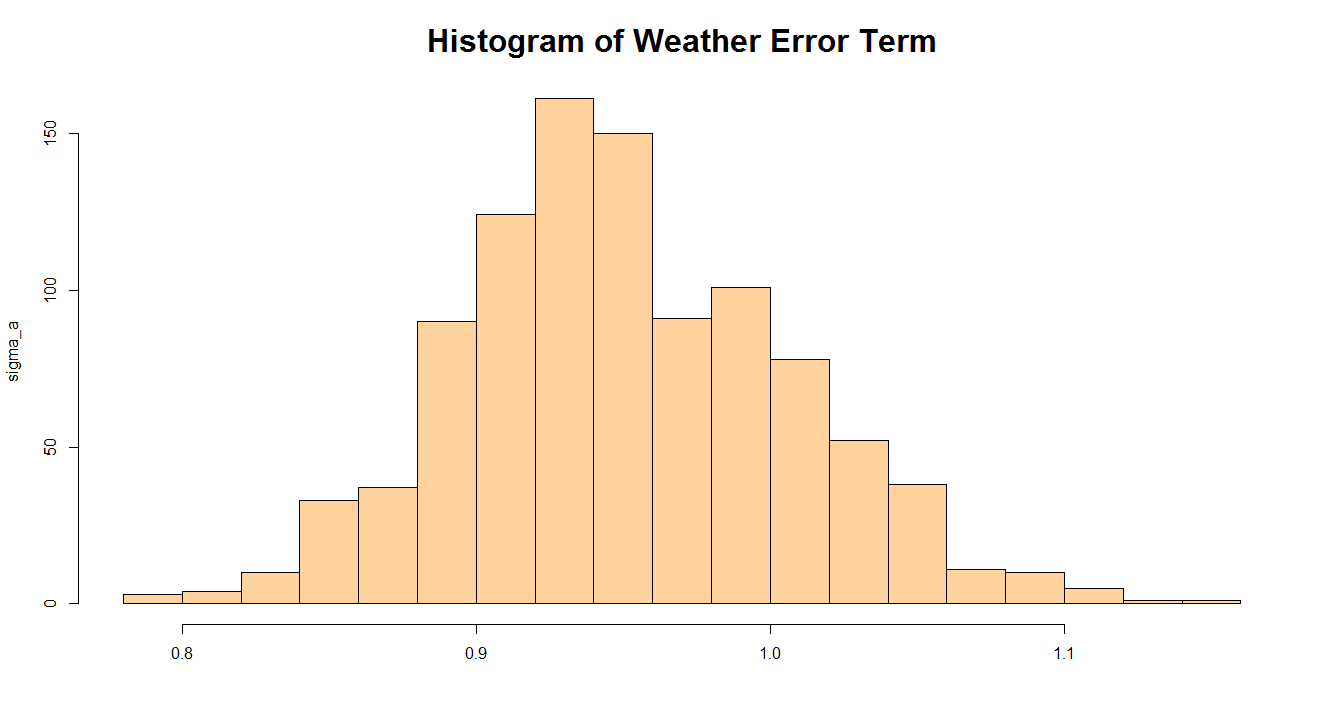}
\includegraphics[width=.45\linewidth,keepaspectratio]{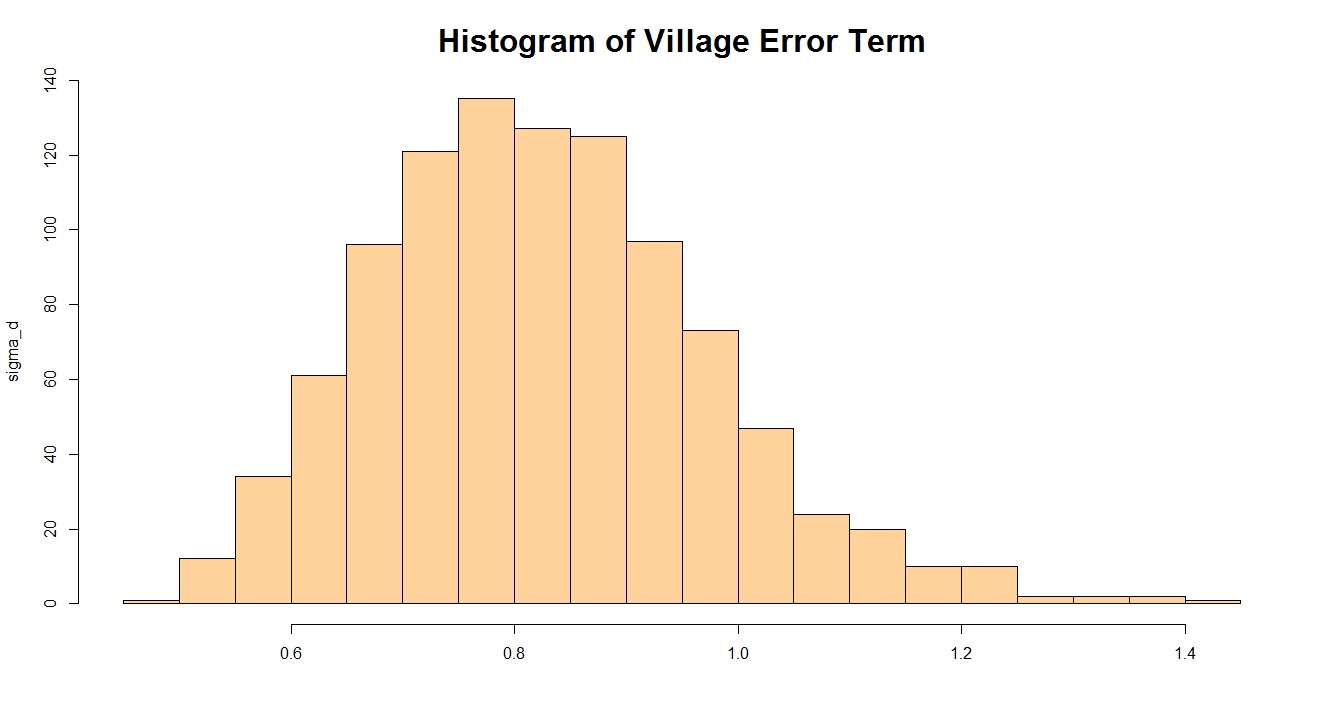}
\includegraphics[width=.45\linewidth,keepaspectratio]{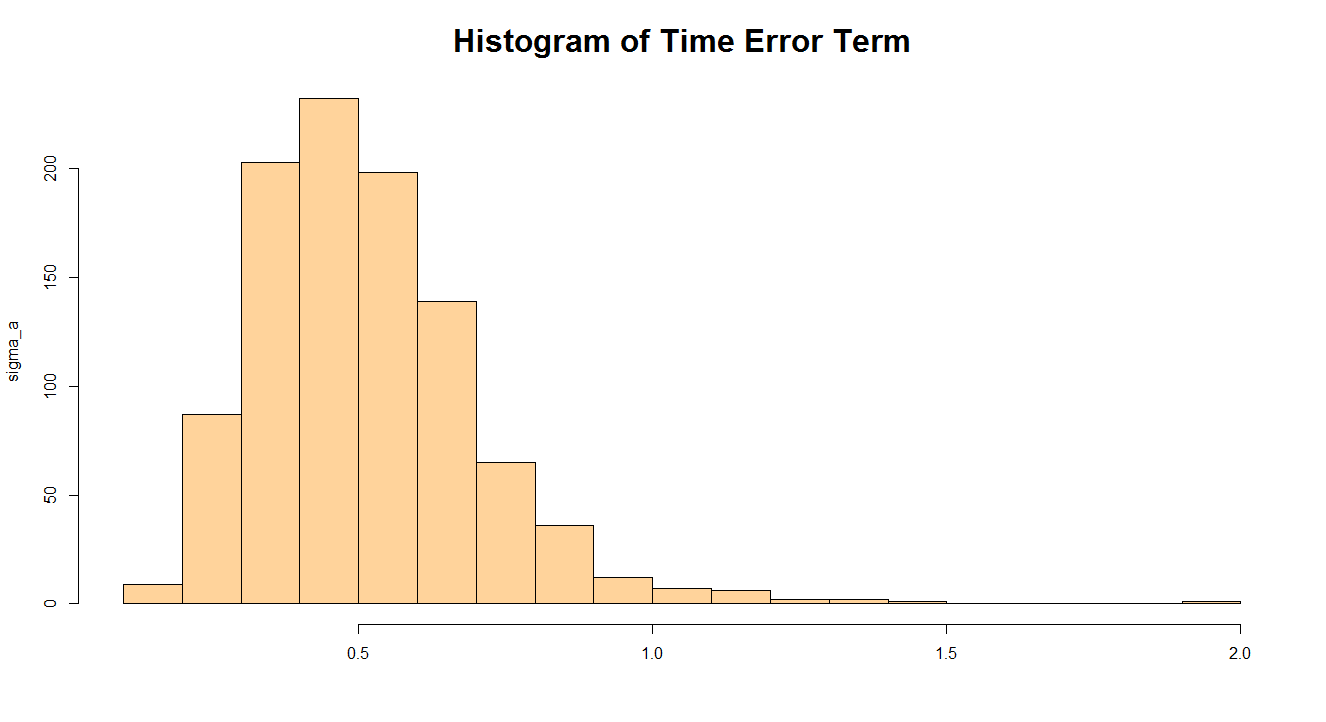}
\includegraphics[width=.45\linewidth,keepaspectratio]{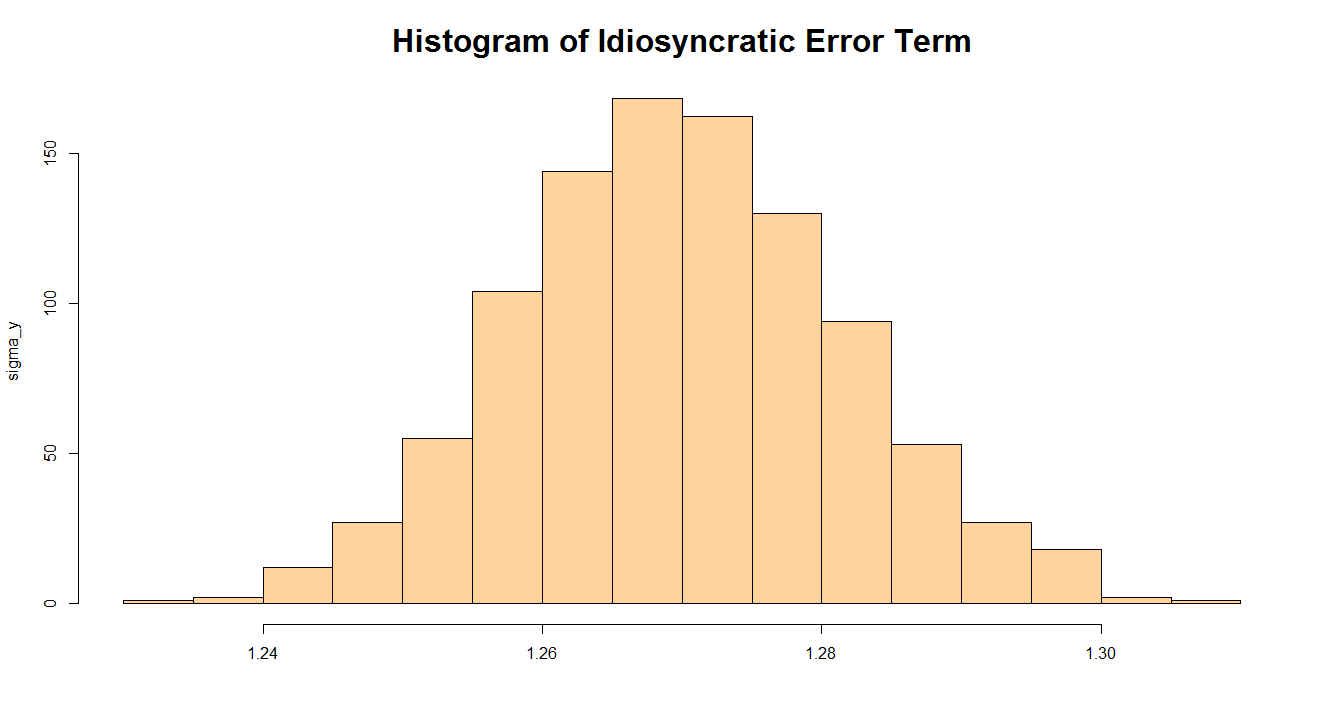}\\
\caption{Histograms of Level Variance from Bayesian Estimation} \label{fig:taui}
\justify \footnotesize \textit{Note}: Histograms are drawn from posterior distributions of variance terms estimated from Column (3) reported in Table~\ref{tab:reg}.
\end{figure}
\end{landscape}

\newpage
\begin{figure}[!htbp]
\includegraphics[width=\textwidth,keepaspectratio]{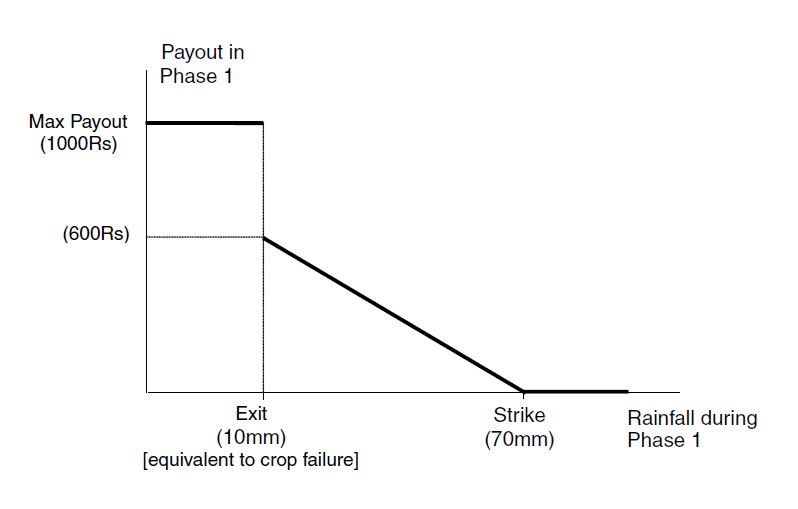}\label{fig:inscon}
\caption{Structure of Insurance Contract}
\footnotesize  \textit{Note}: Figure shows the representative contract structure used in calculating payouts presented in Table~\ref{tab:Ins}. Specifically the figure represents the Phase I contract for the high-payout case. The Phase II contract structure is identical to the Phase I contract with the exception of strike and exit values. The Phase III contract does not differ in structure but has strike and exit values set to insure against excess rainfall, meaning the contract slope is positive. Additionally, the contract is representative of contracts on offer by ICICI Lombard throughout the region of Andhra Pradesh. The contract in each phase has an upper and lower rainfall threshold. The policy pays zero above the strike; otherwise the policy pays a fixed amount for each millimeter of shortfall (excess in the case of Phase III), until the exit is reached. In the contract represented this is 10 Rs. per mm. At the exit point the contract is discontinuous and ``jumps'' to its maximum value. Note that this specific contract comes from \cite{GineEtAl07} and is a slight variation on the contracts presented in \cite{GineEtAl08, GineEtAl12} and \cite{ColeEtAl13}. The contracts presented in these papers share the same structure but vary in the strike and exit values. \\
\textit{Source}: \cite{GineEtAl07}.
\end{figure}

\clearpage
\newpage

\appendix
\section{For Online Publication: Results from Estimation of Crop-Specific Production Functions}\label{sec:appcrop}
In the interests of space we limit the results from the crop-specific regressions presented in the paper and simply present the variance parameter estimates. Here we present the full results from the crop-specific regressions.

For reference, Table~\ref{tab:reg} presents the results from the varying-slope model using OLS, MLE, and Bayesian estimation techniques. Table~\ref{tab:appcrops} presents the MLE and Bayesian point estimates from the production function estimates for each crop.

While point estimates change slightly from the full regression to the crop-by-crop regressions, qualitatively there is no change. For rice, all point estimates in both specifications are positive and significant. For sorghum, all point estimates, save for pesticide, in both specifications are positive and significant. For wheat, all point estimates in both specifications are positive and significant, save for pesticide, which is negative and significant. For maize, all point estimates are positive and significant, save for fertilizer, which is not significant in both specifications. For cotton, all point estimates in both specifications are positive and significant, save for mechanization, which is not significant.

The consistency in point estimates across estimation techniques (OLS, MLE, Bayes) and across divisions of the data demonstrate a remarkably robust set of results.

\section{For Online Publication: Alternative Specifications of the Multilevel Model}\label{sec:approbust}
In order to verify the robustness of our primary results this appendix presents a variety of alternative specifications of our model. The goal is to determine if grouping the data in different ways might have an effect on our inferences.

First, in Table~\ref{tab:appICC}, we test a variety of ``null'' models in which covariates are excluded. In each model we add complexity by including a new level effect. This allows us to see how our estimate of seasonal variability changes as we account for additional level effects. We start in columns (1) and (2) with simply the season, village, and time levels and build up from there. Adding a household level in columns (3) and (4) simply reduces the unassigned variance in the idiosyncratic term and has no effect on the other sources of variance. Adding a parcel level in columns (5) and (6) has a small effect on season-level variance but primarily reduces the amount of variance ascribed to the idiosyncratic error term. In columns (7) and (8) we add in parcel-level measured inputs, which is our preferred specification presented in the paper. Compared to the ``null'' models, the inclusion of inputs has the effect of reduce the variance ascribed to parcel-level effects and the idiosyncratic error term. It also has the effect of increasing the variance due to weather and variance at the village-level. 

Second, we test models which exclude season, village, and temporal levels. Each regression includes the full contingent of measured inputs which we refrain from reporting since our focus is on alternative specifications for the variance structure. The intuition for focusing on the effect of village and time levels in the multilevel model is that our proxy for weather is an interaction effect between village and time. Because of this, the village and/or time levels may be picking up some of the ``true'' effects of weather.

Table~\ref{tab:appalt} presents results from six specifications estimated using MLE.\footnote{The Bayesian estimates tell the same story as the maximum likelihood estimates and so are omitted. Bayesian estimates are available from the authors upon request.} Column (1) reports results from including just a time-level as a proxy for weather. Variation in time alone would be a valid proxy for weather if all locations in the study experienced the same weather. Given the large geographic dispersion it is unsurprising that a weather proxy that disregards local variation in weather accounts for a vary small share of total variance. Column (2) replaces the time-level with a village-level. Variation across villages alone would be a valid proxy for weather if each time period experienced the same weather. Given that there are two distinct seasons in India - wet \emph{Kharif} and dry \emph{Rabi} - and that we examine only five years, this is not unreasonable. We see this in the relatively large share of variance due to the village-level. However, there may be other spatial elements that our village term picks up, such as agronomic suitability or government policy.

Column (3)-(6) include a season-level which is the interaction between village indicators and time indicators. This is our preferred proxy for weather as it allows weather to vary across both villages and time. In column (3) we drop the village and time levels in favor of only the season-level. The result is a relatively large share of variance ascribed to differences across location-time. This might be the ``true'' measure of the effect of weather variability on yield variability. Or, there might be location specific effects and time specific effects that are being captured in this term. To check this, in column (4) we add back in the time-level effect. Having controlled for seasonality, the time effect will only pick up variation in yields across time separate from variation in location-time. Such variation would be from technological change that effects all villages at the same time or, potentially, a nationwide weather event. Given that our data covers only five years the time-only effect is small. In column (5) we add back in the village-level effect instead of the time effect. Having controlled for seasonality, the village effect will only pick up variation in yields across village separate from variation in location-time. Such variation would be from differences in time-invariant agronomic characteristics of the villages or differences in state policies that have different impacts of villages. Given that we have 30 villages across eight Indian states the location-only effect is moderate, relative to other sources of variation. Column (6) is our preferred specification since it controls for time-only, location-only, and location-time events.

Compared to the alternate specifications presented in Table~\ref{tab:appalt}, the specification presented in Column (6) and the paper is a better fit for the data as measured by the log likelihood, Akaike Information Criterion (AIC), Deviance Information Criterion (DIC), and the size of the idiosyncratic residual. The better fit and the ability to control for all potential sources of variance makes this our preferred specification. 

\section{For Online Publication: Details on the Computation of Actuarially Fair Premia and Payout Probabilities}\label{sec:appins}
This appendix provides details on the sources of data for and the computation of actuarially fair premia and payout probabilities. It also contains several robustness checks in which we calculate premia and payout probabilities using data on actual index insurance contracts provided by \cite{GineEtAl07, GineEtAl12} and \cite{ColeEtAl13}.

\subsection{Source of the Data}
Though we have access to daily rainfall data for all 30 villages in the VDSA data set, we limit our calculation of premia and payouts to a subset of 16 villages within the Indian states of Bihar, Jharkhand, Odisha, Andhra Pradesh, and Karnataka. This is done for several reasons.

First, the 16 villages are clustered in nine districts, all of which fall within the semi-arid, tropical wet-and-dry, and humid subtropical climate zones. These zones all share quantitatively similar rainfall patterns and variability during the monsoon season \citep{GineEtAl09}. Despite the shared climate zones, our village rainfall data exhibit a level of variability, due to the wide spatial range, which is of the same order of magnitude as the variability one expects for monsoon rainfall in a single location in the same general region of Central and Eastern India. The actual rainfall values are also representative of what is observed for individual, non-VDSA, locations in the region. In this fashion, our spatial rainfall variability acts as a proxy for temporal variability at the district level. We can therefore view our data as a quasi-panel with multiple observations over space instead of the traditional panel structure of multiple observations over time.

Second, the restriction of analysis to villages from these five states is consistent with the premium data from commercial insurers ICICI-Lombard and IFFCO-Tokio that are most frequently utilized in the academic literature.\footnote{See the various contracts reported in \cite{ClarkeEtAl12a}, \cite{ClarkeEtAl12b}, \cite{ColeEtAl13}, \cite{GineEtAl07}, \cite{GineEtAl08}, and \cite{GineEtAl12}.} Contracts reported therein are largely for various village-weather station pairs in Andhra Pradesh with similar monsoon rainfall characteristics to the villages in the five states we consider from the VDSA.\footnote{\cite{ColeEtAl13} reports contracts from the largest set of village-weather station pairs: Anantapur, Atmakur, Hindupur, Narayanpet, and Mahbubnagar. The three VDSA villages in Andhra Pradesh come from the districts Prakasam and Mahbubnagar.}

Third, inclusion of villages in the Indian states of Madhya Pradesh and Gujarat turn out to be outliers in the humid direction in Phases I and II. By excluding these outliers, and focusing on the subset of 16 villages, we preserve the aforementioned proxy property. This allows us to avoid the use of villages where rainfall-indexed payouts would never occur. While this obviously biases our results, given that we exclude villages that never see a payout, the bias is downwards, making the results presented in the paper conservative in terms of actuarially fair premia and probability of payout for the corresponding
contracts. Had we included these outliers, our conclusions about the excessively high commercial loading factors would be even more extreme.

\subsection{Computation of Actuarially Fair Premia and Payoff Probabilities}
The rain-indexed contracts compute payouts uniformly for all contracts in a given village by using the rainfall values reported from that village's rain gauge. We compute the total rainfall in millimeters for each of the three phases of the monsoon season for that village and insert these phase totals into the contract payout function.\footnote{See the definition of the start and stop of each phase in the text.} We work with payout functions which are characterized by a strike, an exit, and a max payout.\footnote{Again, see the definitions of these terms and of the function in the text.} While there is a variety of contracts on offer these contracts differ not in structure but only in their strike, exit, payout slope, and maximum payout. We calculate actuarially fair premia and payoff probabilities for all five contracts from Andhra Pradesh in \cite{ColeEtAl13} plus the single contract in \cite{GineEtAl07}. While \cite{ClarkeEtAl12a} and \cite{GineEtAl08} present similarly structured contracts, the third phase in these contracts insure against deficit rainfall and therefore are not appropriate for villages in the climate regions where our rainfall data comes from.

We compute actuarially fair premia using standard actuarial principles based on the empirical data. Namely, we evaluate the payout function for each contract at each rainfall phase total and then average these empirical payouts over the 57 data points for each phase of each contract. The total empirical actuarially fair value for each contract is simply the sum of the three corresponding values for each phase of the contract. We compute the payout probability for each contract by averaging the values 1 or 0 corresponding to the occurrence of a payout or not for each of the 171 village-year-phase combinations. This way of accounting keeps track of multiple payouts in each contract year, which is more conservative (providing a higher empirical probability) than if one only looked for whether there was one or more payouts in a given contract year.

\subsection{Actuarially Fair Premia and Payoff Probabilities Under Alternative Contracts}
As explained in Section~\ref{sec:ins}, we presented in the paper only the two most favorable contracts from \cite{ColeEtAl13} plus the contract from \cite{GineEtAl07}. In this appendix we present our calculations from the three less favorable contracts. As can be seen in Table~\ref{tab:appins}, the payout probabilities for these contracts are $8.8$, $4.1$, and $2.9$ percent, making them very unlikely to be taken up by farmers. Additionally, the premia determined by ICICI Lombard are so high in comparison to the actuarially fair prices we have computed empirically, that the corresponding loading factors all exceed three, a sign that these contracts are severely overpriced. Besides our desire to take a conservative approach, another sign that the analysis of these three contracts is not appropriate for use as part of our main argument is that their commercial pricing, while uniformly excessive, is not internally consistent. Indeed, these three contracts come from three weather stations (near the towns of Hindupur, Narayanpet, and Mahbubnagar) which are all in the same climate zone, are all geographically near each other, and all share the same maximum payout and payout slope. Therefore, since they only differ by strike and exit values, actuarial principles dictate that the most favorable strikes and exits should have the highest premium. That would imply that the premium for the Hindupur contract should be lower than for the other two contracts. In reality, as seen in \cite{ColeEtAl13}, it is higher. When we consider these contracts, in combination with our preceding analysis, we find it unsurprising that \cite{ColeEtAl13} found little uptake of weather index insurance contracts in their randomized control trial, except when farmers were offered a cash discount. Our findings provide a simple explanation for the experimental results: the contracts hold little value for rural households.

\newpage
\setcounter{table}{0}
\renewcommand{\thetable}{A\arabic{table}}

\begin{landscape}
	\begin{table}[!htbp] \centering 
		\caption{Classical and Bayesian Estimation of Production Functions by Crop Type} \label{tab:appcrops} 
		\scalebox{.58}
		{\setlength{\linewidth}{.1cm}\newcommand{\contents}
			{\begin{tabular}{@{\extracolsep{5pt}}lD{.}{.}{-3} D{.}{.}{-3}  D{.}{.}{-3} D{.}{.}{-3} D{.}{.}{-1} D{.}{.}{-3}  D{.}{.}{-3} D{.}{.}{-3} D{.}{.}{-1}  D{.}{.}{-3} D{.}{.}{-3} D{.}{.}{-3}} 
					\\[-1.8ex]\hline 
					\hline \\[-1.8ex] 
					& \multicolumn{2}{c}{Rice}		& \multicolumn{2}{c}{Sorghum}	& \multicolumn{2}{c}{Wheat}		& \multicolumn{2}{c}{Maize}		& \multicolumn{2}{c}{Cotton}	\\ 
					& \multicolumn{1}{c}{\textit{MLE}} &	\multicolumn{1}{c}{\textit{Bayes}} & \multicolumn{1}{c}{\textit{MLE}} &	\multicolumn{1}{c}{\textit{Bayes}} & \multicolumn{1}{c}{\textit{MLE}} &	\multicolumn{1}{c}{\textit{Bayes}} & \multicolumn{1}{c}{\textit{MLE}} &	\multicolumn{1}{c}{\textit{Bayes}} & \multicolumn{1}{c}{\textit{MLE}} &	\multicolumn{1}{c}{\textit{Bayes}} \\
					& \multicolumn{1}{c}{(1)} &	\multicolumn{1}{c}{(2)} & \multicolumn{1}{c}{(3)} &	\multicolumn{1}{c}{(4)} & \multicolumn{1}{c}{(5)} &	\multicolumn{1}{c}{(6)} & \multicolumn{1}{c}{(7)} &	\multicolumn{1}{c}{(8)} & \multicolumn{1}{c}{(9)} &	\multicolumn{1}{c}{(10)} \\
					\midrule
					log labor 			& 1.549^{***}	& 1.552^{***}	& 0.472^{***}	& 0.473^{***}	& 1.411^{***}	& 1.411^{***}	& 0.869^{***}	& 0.871^{***}	& 1.196^{***}	& 1.201^{***}	\\ 
					& (0.054)		& (0.054)		& (0.039)		& (0.041)		& (0.071)		& (0.070)		& (0.073)		& (0.077)		& (0.080)		& (0.081)		\\ 
					log fertilizer 		& 0.594^{***}	& 0.594^{***}	& 0.123^{***}	& 0.123^{***}	& 0.120^{***}	& 0.120^{***}	& -0.025		& -0.026		& 0.249^{***}	& 0.248^{***}	\\
					& (0.026)		& (0.027)		& (0.025)		& (0.026)		& (0.027)		& (0.028)		& (0.030)		& (0.032)		& (0.037)		& (0.039)		\\ 
					log mechanization	& 0.196^{***}	& 0.197^{***}	& 0.232^{***}	& 0.232^{***}	& 0.449^{***}	& 0.450^{***}	& 0.042^{***}	& 0.041^{***}	& -0.001		& -0.003		\\ 
					& (0.013)		& (0.013)		& (0.019)		& (0.020)		& (0.024)		& (0.024)		& (0.014)		& (0.015)		& (0.023)		& (0.023)		\\ 
					log pesticides 		& 0.114^{***}	& 0.114^{***}	& 0.012			& 0.011			& -0.161^{***}	& -0.162^{***}	& 0.045^{*}		& 0.045^{***}	& 0.140^{***}	& 0.138^{***}	\\ 
					& (0.014)		& (0.014)		& (0.011)		& (0.011)		& (0.035)		& (0.035)		& (0.025)		& (0.024)		& (0.020)		& (0.020)		\\ 
					\midrule
					observations		& \multicolumn{2}{c}{5,572}		& \multicolumn{2}{c}{2,720}		& \multicolumn{2}{c}{1,625}		& \multicolumn{2}{c}{1,073}		& \multicolumn{2}{c}{952}		\\
					num. seasons		& \multicolumn{2}{c}{10}		& \multicolumn{2}{c}{5}			& \multicolumn{2}{c}{11}		& \multicolumn{2}{c}{10}		& \multicolumn{2}{c}{5}			\\
					Log Likelihood 		& \multicolumn{1}{c}{-11,264}	& & \multicolumn{1}{c}{-3,507} 	& & \multicolumn{1}{c}{-3,183} 	& & \multicolumn{1}{c}{-1,521} 	& & \multicolumn{1}{c}{-1,308} 	& \\
					Akaike Inf. Crit. 	& \multicolumn{1}{c}{22,551}	& & \multicolumn{1}{c}{7,036} 	& & \multicolumn{1}{c}{6,388} 	& & \multicolumn{1}{c}{3,064} 	& & \multicolumn{1}{c}{2,639} 	& \\
					Deviance Inf. Crit.	& \multicolumn{1}{c}{22,482}	& \multicolumn{1}{c}{21,185} 	& \multicolumn{1}{c}{6,961} 	& \multicolumn{1}{c}{6,872} 	& \multicolumn{1}{c}{6,328} 	& \multicolumn{1}{c}{5,979} & \multicolumn{1}{c}{2,996} & \multicolumn{1}{c}{2,721} & \multicolumn{1}{c}{2,574} & \multicolumn{1}{c}{2,532} \\
					\hline 
					\hline \\[-1.8ex] 
					\multicolumn{11}{p{\linewidth}}{\footnotesize \textit{Note}: Dependent variable is log of yield. All specifications include a statistically significant intercept term. Columns (1) and (2) are estimates of the production function for rice. Columns (3) and (4) are estimates of the production function for sorghum. Columns (5) and (6) are estimates of the production function for wheat. Columns (7) and (8) are estimates of the production function for maize. Columns (9) and (10) are estimates of the production function for cotton. Odd numbered columns are maximum likelihood estimates while even numbered columns are Bayesian estimates. Bayesian calculations use a burn-in period of 5,000 iterations and an additional 5,000 iterations to ensure convergence. Standard errors are reported in parentheses ($^{*}$p$<$0.1; $^{**}$p$<$0.05; $^{***}$p$<$0.01).} \\ 
				\end{tabular}}
				\setbox0=\hbox{\contents}
				\setlength{\linewidth}{\wd0-2\tabcolsep-.25em}
				\contents}
		\end{table}
	\end{landscape}
	
	\begin{landscape}
		\begin{table}[!htbp] \centering 
			\caption{Estimated Variance, ICCs, and Variance Shares from Null Multilevel Regression Models} \label{tab:appICC} 
			\scalebox{.7}
			{\setlength{\linewidth}{.1cm}\newcommand{\contents}
				{\begin{tabular}{@{\extracolsep{5pt}}lD{.}{.}{-3} D{.}{.}{-3} D{.}{.}{-3} D{.}{.}{-3}  D{.}{.}{-3} D{.}{.}{-3} D{.}{.}{-3} D{.}{.}{-3}} 
						\\[-1.8ex]\hline 
						\hline \\[-1.8ex] 
						& \multicolumn{1}{c}{\textit{MLE}} &	\multicolumn{1}{c}{\textit{Bayes}} & \multicolumn{1}{c}{\textit{MLE}} &	\multicolumn{1}{c}{\textit{Bayes}} & \multicolumn{1}{c}{\textit{MLE}} &	\multicolumn{1}{c}{\textit{Bayes}} & \multicolumn{1}{c}{\textit{MLE}} &	\multicolumn{1}{c}{\textit{Bayes}}  \\
						& \multicolumn{1}{c}{(1)} &	\multicolumn{1}{c}{(2)} & \multicolumn{1}{c}{(3)} &	\multicolumn{1}{c}{(4)} & \multicolumn{1}{c}{(5)} &	\multicolumn{1}{c}{(6)} & \multicolumn{1}{c}{(7)} &	\multicolumn{1}{c}{(8)}  \\
						\midrule
						& & & & & & & & \\
						\multicolumn{9}{c}{Panel A: \textit{Variance Parameter Estimates}}	\\  
						& & & & & & & & \\
						parcel $(\sigma_{1}^2)$			& 		& 		& 		& 		& 2.226 & 2.301	& 0.933 & 1.098	\\
						household $(\sigma_{2}^2)$		& 		& 		& 0.030 & 0.020	& 0.000 & 0.002	& 0.002 & 0.004	\\
						season $(\sigma_{3}^2)$			& 0.557	& 0.567	& 0.557 & 0.564	& 0.479 & 0.487	& 0.790 & 0.903	\\
						village $(\sigma_{4}^2)$		& 0.166	& 0.339	& 0.165 & 0.335	& 0.258 & 0.402	& 0.623 & 0.682	\\
						time $(\sigma_{5}^2)$			& 0.026	& 0.100	& 0.026 & 0.112	& 0.035 & 0.095	& 0.126 & 0.261	\\
						idiosyncratic $(\sigma^2)$		& 3.853	& 3.920	& 3.826 & 3.901	& 2.068 & 2.082	& 1.623 & 1.613	\\
						\midrule
						& & & & & & & & \\
						\multicolumn{9}{c}{Panel B: \textit{Intraclass Correlation Coefficients}}	\\ 
						& & & & & & & & \\
						parcel							&		&  		& 		& 		& 0.439	& 0.429	& 0.228	& 0.241	\\
						household						&		&  		& 0.007	& 0.004	& 0.439	& 0.429	& 0.228	& 0.242	\\
						season 							& 0.121	& 0.115	& 0.127 & 0.118	& 0.534 & 0.520	& 0.421 & 0.440	\\
						village 						& 0.157	& 0.184	& 0.163 & 0.186	& 0.585 & 0.594	& 0.573 & 0.589	\\
						time 							& 0.163	& 0.204	& 0.169 & 0.209	& 0.592 & 0.612	& 0.604 & 0.646	\\
						\midrule
						& & & & & & & & \\
						\multicolumn{9}{c}{Panel C: \textit{Shares of Variance From Each Level}}	\\ 
						& & & & & & & & \\
						parcel			&							&							& 							&							& \multicolumn{1}{c}{44\%}	& \multicolumn{1}{c}{43\%}	& \multicolumn{1}{c}{23\%}	& \multicolumn{1}{c}{24\%}	\\
						household		&							&							& \multicolumn{1}{c}{01\%}	& \multicolumn{1}{c}{00\%}	& \multicolumn{1}{c}{00\%}	& \multicolumn{1}{c}{00\%}	& \multicolumn{1}{c}{00\%}	& \multicolumn{1}{c}{00\%}	\\
						season			& \multicolumn{1}{c}{12\%}	& \multicolumn{1}{c}{12\%}	& \multicolumn{1}{c}{12\%}	& \multicolumn{1}{c}{12\%}	& \multicolumn{1}{c}{09\%}	& \multicolumn{1}{c}{09\%}	& \multicolumn{1}{c}{19\%}	& \multicolumn{1}{c}{20\%}	\\
						village			& \multicolumn{1}{c}{04\%}	& \multicolumn{1}{c}{07\%}	& \multicolumn{1}{c}{04\%}	& \multicolumn{1}{c}{07\%}	& \multicolumn{1}{c}{05\%}	& \multicolumn{1}{c}{07\%}	& \multicolumn{1}{c}{15\%}	& \multicolumn{1}{c}{15\%}	\\
						time			& \multicolumn{1}{c}{01\%}	& \multicolumn{1}{c}{02\%}	& \multicolumn{1}{c}{01\%}	& \multicolumn{1}{c}{02\%}	& \multicolumn{1}{c}{01\%}	& \multicolumn{1}{c}{02\%}	& \multicolumn{1}{c}{03\%}	& \multicolumn{1}{c}{06\%}	\\
						idiosyncratic	& \multicolumn{1}{c}{84\%}	& \multicolumn{1}{c}{80\%}	& \multicolumn{1}{c}{83\%}	& \multicolumn{1}{c}{79\%}	& \multicolumn{1}{c}{41\%}	& \multicolumn{1}{c}{39\%}	& \multicolumn{1}{c}{40\%}	& \multicolumn{1}{c}{35\%}	\\
						\midrule
						Observations & \multicolumn{1}{c}{11,942} & \multicolumn{1}{c}{11,942} & \multicolumn{1}{c}{11,942} & \multicolumn{1}{c}{11,942} & \multicolumn{1}{c}{11,942} & \multicolumn{1}{c}{11,942} & \multicolumn{1}{c}{11,942} & \multicolumn{1}{c}{11,942} \\ 
						Log Likelihood  & \multicolumn{1}{c}{-25,231} &	& \multicolumn{1}{c}{-25,226} &	& \multicolumn{1}{c}{-24,422} &	& \multicolumn{1}{c}{-22,268} &	\\ 
						Akaike Inf. Crit. & \multicolumn{1}{c}{50,480} & & \multicolumn{1}{c}{50,473} & & \multicolumn{1}{c}{48,867} & & \multicolumn{1}{c}{44,598} & \\ 
						Deviance Inf. Crit. & \multicolumn{1}{c}{50,435} &  \multicolumn{1}{c}{50,384} & \multicolumn{1}{c}{50,426} & \multicolumn{1}{c}{50,378} & \multicolumn{1}{c}{48,818} & \multicolumn{1}{c}{46,115} & \multicolumn{1}{c}{44,828}	& \multicolumn{1}{c}{42,578} \\ 
						\hline 
						\hline \\[-1.8ex] 
						\multicolumn{9}{p{\linewidth}}{\footnotesize \textit{Note}: Estimates of the variance parameters in Panel A, Columns (1)-(6), come from estimations of regressions without covariates (so-called ``null'' models) that have been estimated at various levels. Estimates of the variance parameters in Panel A, Columns (7) and (8) come from estimation of models reported in Table~\ref{tab:reg}. Columns (1) and (2) contain no covariates and data clustered only at the season-,village-, and time-level. Column (3) and (4) contain no covariates and add a cluster at the household-level. Column (5) and (6) contain no covariates and add a cluster at the parcel-level. Column (7) and (8) contain covariates and data clustered at all levels. Odd numbered columns are maximum likelihood estimates while even numbered columns are Bayesian estimates. Bayesian calculations use a burn-in period of 5,000 iterations and an additional 5,000 iterations to ensure convergence. Variances $\sigma_{1}^2$, $\sigma_{2}^2$, $\sigma_{3}^2$, $\sigma_{4}^2$, $\sigma_{5}^2$ represent the variance in crop yield that comes from the corresponding level. The final variance parameter $(\sigma^2)$ corresponds to the idiosyncratic or unexplained portion of the model. Intraclass correlation coefficients in Panel B are calculated using the formulas in equations~\eqref{eq:icc1}-\eqref{eq:icc5}. Panel C decomposes the ICC into percent of variance accorded to each level.} \\ 
					\end{tabular}}
					\setbox0=\hbox{\contents}
					\setlength{\linewidth}{\wd0-2\tabcolsep-.25em}
					\contents}
			\end{table} 
		\end{landscape}
		
		\begin{table}[!htbp] \centering 
			\caption{Estimated Variance, ICCs, Variance Shares from Alternative Specifications} \label{tab:appalt} 
			\scalebox{.9}
			{\setlength{\linewidth}{.1cm}\newcommand{\contents}
				{\begin{tabular}{@{\extracolsep{5pt}}lD{.}{.}{-3} D{.}{.}{-3}  D{.}{.}{-3} D{.}{.}{-3} D{.}{.}{-1} D{.}{.}{-3}   } 
						\\[-1.8ex]\hline 
						\hline \\[-1.8ex] 
						& \multicolumn{1}{c}{(1)} &	\multicolumn{1}{c}{(2)} & \multicolumn{1}{c}{(3)} &	\multicolumn{1}{c}{(4)} & \multicolumn{1}{c}{(5)} &	\multicolumn{1}{c}{(6)}  \\
						\midrule
						& & & & & &  \\
						\multicolumn{7}{c}{Panel A: \textit{Variance Parameter Estimates}}	\\ 
						& & & & & &   \\
						parcel 			& 1.378 & 1.171	& 1.000 & 1.000	& 0.992 & 0.933		\\
						household 		& 0.179	& 0.000	& 0.002 & 0.002	& 0.002 & 0.002		\\
						season			& 		& 		& 1.343	& 1.290	& 0.912 & 0.790		\\
						village			& 		& 0.726	& 		& 		& 0.548 & 0.623		\\					
						time			& 0.146	& 		& 		& 0.031	& 		& 0.126		\\
						idiosyncratic	& 1.957	& 2.059	& 1.623 & 1.623	& 1.626 & 1.623		\\
						\midrule
						& & & & & &  \\
						\multicolumn{7}{c}{Panel B: \textit{Intraclass Correlation Coefficients}}	\\ 
						& & & & & & \\
						parcel			& 0.377	& 0.296	& 0.252 & 0.251	& 0.243 & 0.228		\\
						household		& 0.425	& 0.296	& 0.252 & 0.252	& 0.244 & 0.228		\\
						season			& 		& 		& 0.591	& 0.577	& 0.467 & 0.421		\\
						village			& 		& 0.479	& 		& 		& 0.601	& 0.573		\\					
						time			& 0.465	& 		& 		& 0.592	& 		& 0.604		\\
						\midrule
						& & & & & &  \\
						\multicolumn{7}{c}{Panel C: \textit{Shares of Variance From Each Level}}	\\ 
						& & & & & &  \\
						parcel			& \multicolumn{1}{c}{38\%}		& \multicolumn{1}{c}{30\%}		& \multicolumn{1}{c}{25\%}		& \multicolumn{1}{c}{25\%}		& \multicolumn{1}{c}{24\%} & \multicolumn{1}{c}{23\%}				\\
						household		& \multicolumn{1}{c}{05\%}		& \multicolumn{1}{c}{00\%}		& \multicolumn{1}{c}{00\%}		& \multicolumn{1}{c}{00\%}		& \multicolumn{1}{c}{00\%} & \multicolumn{1}{c}{00\%}			\\
						season			& \multicolumn{1}{c}{}			& \multicolumn{1}{c}{}			& \multicolumn{1}{c}{34\%}		& \multicolumn{1}{c}{32\%}		& \multicolumn{1}{c}{22\%} & \multicolumn{1}{c}{19\%}			\\
						village			& \multicolumn{1}{c}{}			& \multicolumn{1}{c}{18\%}		& \multicolumn{1}{c}{}			& \multicolumn{1}{c}{}			& \multicolumn{1}{c}{13\%} & \multicolumn{1}{c}{15\%}			\\
						time			& \multicolumn{1}{c}{04\%}		& \multicolumn{1}{c}{}			& \multicolumn{1}{c}{}			& \multicolumn{1}{c}{02\%}		& \multicolumn{1}{c}{}	   & \multicolumn{1}{c}{03\%}			\\
						idiosyncratic	& \multicolumn{1}{c}{53\%}		& \multicolumn{1}{c}{52\%}		& \multicolumn{1}{c}{41\%}		& \multicolumn{1}{c}{41\%}		& \multicolumn{1}{c}{40\%} & \multicolumn{1}{c}{40\%}			\\
						\midrule
						observations & \multicolumn{1}{c}{11,942} & \multicolumn{1}{c}{11,942} & \multicolumn{1}{c}{11,942} & \multicolumn{1}{c}{11,942} & \multicolumn{1}{c}{11,942} & \multicolumn{1}{c}{11,942} \\ 
						log likelihood & \multicolumn{1}{c}{-23,423} & \multicolumn{1}{c}{-23,323} & \multicolumn{1}{c}{-22,296} & \multicolumn{1}{c}{-22,294} & \multicolumn{1}{c}{-22,274} & \multicolumn{1}{c}{-22,268} \\ 
						Akaike Inf. Crit. & \multicolumn{1}{c}{46,904} & \multicolumn{1}{c}{46,704} & \multicolumn{1}{c}{44,650} & \multicolumn{1}{c}{44,649} & \multicolumn{1}{c}{44,608} & \multicolumn{1}{c}{44,598} \\ 
						Deviance Inf. Crit. & \multicolumn{1}{c}{46,599} & \multicolumn{1}{c}{46,400} & \multicolumn{1}{c}{44,351} & \multicolumn{1}{c}{44,350} & \multicolumn{1}{c}{44,310} & \multicolumn{1}{c}{44,828} \\ 
						\hline 
						\hline \\[-1.8ex] 
						\multicolumn{7}{p{\linewidth}}{\footnotesize \textit{Note}: Estimates of the variance parameters in Panel A come from alternative specifications of the models presented in Table~\ref{tab:reg}. All models contain crop-specific covariates and crop-specific intercepts. All six specifications are estimated using MLE. Column (1) reports results using only a time-level to proxy for weather while Column (2) uses only a village-level to proxy for weather. Column (3) drops the village and time levels in favor of the season-level. Column (4) adds back in the time-level while Column (5) adds back in the village-level. Column (6) includes season, village, and time levels. Intraclass correlation coefficients in Panel B are calculated using the formulas in equations~\eqref{eq:icc1}-\eqref{eq:icc5}.  Panel C decomposes the ICC into percent of variance accorded to each level.} \\ 
					\end{tabular}}
					\setbox0=\hbox{\contents}
					\setlength{\linewidth}{\wd0-2\tabcolsep-.25em}
					\contents}
			\end{table}
			
			\begin{table}[!htbp] \centering 
				\caption{Actuarial Values and Payout Probabilities for Alternative Contracts} \label{tab:appins} 
				\scalebox{.8}
				{\setlength{\linewidth}{.1cm}\newcommand{\contents}
					{\begin{tabular}{@{\extracolsep{1pt}}lD{.}{.}{-3}D{.}{.}{-3} D{.}{.}{-3} D{.}{.}{-3} D{.}{.}{-3} D{.}{.}{-3} D{.}{.}{-3}}
							\\[-1.8ex]\hline 
							\hline \\[-1.8ex]
							& \multicolumn{3}{c}{\textit{Contract Structure}} & \multicolumn{4}{c}{\textit{Summary Statistics}}  \\ 
							&  &  &  & \multicolumn{1}{c}{Actuarially Fair} & \multicolumn{1}{c}{Probability} & \multicolumn{1}{c}{Loading} & \multicolumn{1}{c}{Years until} \\ 
							& \multicolumn{1}{c}{Strike} & \multicolumn{1}{c}{Exit} & \multicolumn{1}{c}{Max Payout} & \multicolumn{1}{c}{Premium} & \multicolumn{1}{c}{of Payout} & \multicolumn{1}{c}{Factor} & \multicolumn{1}{c}{Payout} \\ 
							\midrule
							& & & & & & & \\
							\multicolumn{8}{c}{Panel A}	\\ 
							\multicolumn{8}{c}{\textit{High-Payout Contract Structure}}	\\ 
							& & & & & & & \\
							\multicolumn{1}{l}{Phase I}		& 45	& 5		& 1000	& \multirow{3}{*}{63.67}	& \multirow{3}{*}{8.77\%}  & \multirow{3}{*}{4.55}	& \multirow{3}{*}{3.80}  \\ 
							\multicolumn{1}{l}{Phase II}	& 55	& 5		& 1000	&	&  &  &  \\ 
							\multicolumn{1}{l}{Phase III} 	& 500	& 570	& 1000	&  	&  &  &  \\ 
							\midrule
							& & & & & & & \\
							\multicolumn{8}{c}{Panel B}	\\ 
							\multicolumn{8}{c}{\textit{Medium-Payout Contract Structure}}	\\ 
							& & & & & & & \\
							\multicolumn{1}{l}{Phase I}		& 25	& 0			& 1000	& \multirow{3}{*}{81.71}	& \multirow{3}{*}{4.14\%}  & \multirow{3}{*}{3.73}	& \multirow{3}{*}{8.06}  \\ 
							\multicolumn{1}{l}{Phase II}	& 15	& 0			& 1000	&  &  &  &  \\ 
							\multicolumn{1}{l}{Phase III}	& 500	& 580		& 1000	&  &  &  &  \\ 
							\midrule
							& & & & & & & \\
							\multicolumn{8}{c}{Panel C}	\\ 
							\multicolumn{8}{c}{\textit{Low-Payout Contract Structure}}	\\ 
							& & & & & & & \\
							\multicolumn{1}{l}{Phase I}		& 30	& 5			& 1000	& \multirow{3}{*}{42.02}	& \multirow{3}{*}{2.92\%} & \multirow{3}{*}{8.33}	& \multirow{3}{*}{11.40}   \\ 
							\multicolumn{1}{l}{Phase II}	& 30	& 5			& 1000	&  &  &  &  \\ 
							\multicolumn{1}{l}{Phase III}	& 500	& 575 		& 1000 	&  &  &  &  \\ 
							\\[-1.8ex]\hline 
							\hline \\[-1.8ex] 
							\multicolumn{8}{p{\linewidth}}{\footnotesize \textit{Note}: All contracts come from \cite{ColeEtAl13}. While each contract is designed for a specific village/weather station, we follow \cite{GineEtAl07} in utilizing a large representative set of rainfall data to calculate actuarially fair premia and the probability of payout. Actuarially fair premia are calculated using standard actuarial principles. Payout probability is the average occurrence of a payout. Loading factors are calculated as $1 + \lambda = \frac{\textrm{paid price}}{\textrm{actuarially fair price}}$. Years until payout are calculated by dividing the inverse of the probability of payout by three, which assumes payout events occur as a Poisson process.} \\ 
						\end{tabular}}
						\setbox0=\hbox{\contents}
						\setlength{\linewidth}{\wd0-2\tabcolsep-.25em}
						\contents}
				\end{table}	
\end{document}